\documentclass[fleqn,usenatbib]{mnras}

\usepackage{newtxtext,newtxmath}

\usepackage[T1]{fontenc}
\usepackage{ae,aecompl}

\usepackage{hyperref}
\usepackage[nameinlink]{cleveref}
\usepackage[linesnumbered]{algorithm2e}

\usepackage{graphicx}	
\usepackage{amsmath}	
\usepackage{amssymb}	
\newcommand{\bjdtdb}{BJD$_\textrm{TDB}\ $}      
\newcommand{\kms}{$\,$km$\,$s$^{-1}$}

\newcommand{\msun}{\ensuremath{\,M_\odot}}

\title[The PDS\,110 observing campaign]{The PDS\,110 observing campaign -- photometric and spectroscopic observations reveal eclipses are aperiodic\thanks{Based on observations collected with Las Cumbres Observatory under program LCO2017AB-003, and the European Southern Observatory, Chile under programme 299.C-5047}}

\author[H.P. Osborn et al.]{
\parbox{\textwidth}{
H.P.~Osborn\thanks{E-mail: \href{hugh.osborn@lam.fr}{hugh.osborn@lam.fr}}$^{\rm 1}$,
M.~Kenworthy$^{\rm 2}$,
J.E.~Rodriguez$^{\rm 3}$,
E.J.W.~de~Mooij$^{4}$,
G.M.~Kennedy$^{\rm 5, 6}$,
H.~Relles$^{7}$,
E.~Gomez,$^{7}$,
M.~Hippke$^{8}$,
M.~Banfi$^{\ddagger}$,
L.~Barbieri$^{\ddagger}$,
I.S.~Becker$^{9}$,
P.~Benni$^{\ddagger \rm{,10}}$,
P.~Berlind$^{3}$,
A.~Bieryla$^{3}$,
G.~Bonnoli$^{\rm{11}}$,
H.~Boussier$^{\ddagger}$,
S.M.~Brincat$^{\ddagger}$,
J.~Briol$^{\ddagger}$,
M.R.~Burleigh$^{\rm 12}$,
T.~Butterley$^{\rm 13}$,
M.L.~Calkins$^{\rm 3}$,
P.~Chote$^{\rm 5}$,
S.~Ciceri$^{\rm 14}$,
M.~Deldem$^{\ddagger}$,
V.S.~Dhillon$^{\rm 15, 16}$,
E.~Dose$^{\ddagger}$,
F.~Dubois$^{\ddagger \rm{, 17}}$,
S.~Dvorak$^{\ddagger}$,
G.A.~Esquerdo$^{\rm 3}$,
D.F.~Evans$^{\rm 18}$,
S.~Ferratfiat$^{\ddagger}$,
S.J.~Fossey$^{\rm 19, 20}$,
M.N.~G{\"u}nther$^{\rm 21}$,
J.~Hall$^{\ddagger}$,
F.-J.~Hambsch$^{\ddagger \rm{, 22}}$,
E.~Herrero$^{\rm 23}$,
K.~Hills$^{\ddagger}$,
R.~James$^{\ddagger}$,
R. Jayawardhana$^{\rm 24}$,
S.~Kafka$^{\rm 25}$,
T.L.~Killestein$^{\ddagger \rm{, 4}}$,
C.~Kotnik$^{\ddagger}$,
D.W.~Latham$^{\rm 3}$,
D.~Lemay$^{\ddagger}$,
P.~Lewin$^{\ddagger}$,
S.~Littlefair$^{\rm 15}$,
C.~Lopresti$^{\ddagger}$,
M.~Mallonn$^{\rm 26}$,
L.~Mancini$^{\rm 27,\,28,\,29}$,
A.~Marchini$^{\ddagger \rm{, 11}}$,
J.J.~McCormac$^{\rm 5,6}$,
G.~Murawski$^{\rm \ddagger ,30}$,
G.~Myers$^{\ddagger \,25}$,
R.~Papini$^{\ddagger}$,
V.~Popov$^{\ddagger, 31}$,
U.~Quadri$^{\ddagger}$,
S.N.~Quinn$^{\rm 3}$,
L.~Raynard$^{\rm 12}$,
L.~Rizzuti$^{\ddagger}$,
J.~Robertson$^{\rm 32}$,
F.~Salvaggio$^{\ddagger}$,
A.~Scholz$^{\rm 33}$,
R.~Sfair$^{\rm 9}$,
A.~M.~S.~Smith$^{\rm 34}$,
J.~Southworth$^{\rm 18}$,
T.G.~Tan$^{\rm \ddagger, 35}$
S.~Vanaverbeke$^{\rm \ddagger, 17}$, 
E.O.~Waagen$^{\rm 23}$,
C.A.~Watson$^{\rm 36}$,
R.G.~West$^{\rm 5, 6}$,
O.C.~Winter$^{\rm 9}$,
P.J.~Wheatley$^{\rm 5, 6}$,
R.W.~Wilson$^{\rm 13}$,
G.~Zhou$^{\rm 3}$
\\
}
\\
Affiliations are listed at the end of the paper.
}
\date{Accepted XXX. Received YYY; in original form ZZZ}

\pubyear{2019}

\begin{document}
\label{firstpage}
\pagerange{\pageref{firstpage}--\pageref{lastpage}}

\maketitle

\begin{abstract}
PDS\,110 is a young disk-hosting star in the Orion OB1A association.
Two dimming events of similar depth and duration were seen in 2008 (WASP) and 2011 (KELT), consistent with an object in a closed periodic orbit.
In this paper we present data from a ground-based observing campaign designed to measure the star both photometrically and spectroscopically during the time of predicted eclipse in September 2017.
Despite high-quality photometry, the predicted eclipse did not occur, although coherent structure is present suggesting variable amounts of stellar flux or dust obscuration.
We also searched for RV oscillations caused by any hypothetical companion and can rule out close binaries to 0.1 $M_s$.
A search of Sonneberg plate archive data also enabled us to extend the photometric baseline of this star back more than 50 years, and similarly does not re-detect any deep eclipses.
Taken together, they suggest that the eclipses seen in WASP and KELT photometry were due to aperiodic events.
It would seem that PDS\,110 undergoes stochastic dimmings that are shallower and shorter-duration than those of UX Ori variables, but may have a similar mechanism.
\end{abstract}




\section{Introduction}

In the process of planet formation, a circumstellar disk is formed around a star.
This circumstellar disk, and the subsequent formation of protoplanetary cores, can be probed and studied by direct imaging, but also through photometric observations of young stars.
Protoplanetary cores subsequently draw matter from the circumstellar disk, potentially forming a circumplanetary disk that fills a significant fraction of the Hill sphere of the planet \citep[e.g. see reviews by][]{Armitage11,Kley12}, which accretes either onto the exoplanet, into exo-moons, or possibly exo-rings \citep{Canup02,Magni04,Ward10}.
Such objects can also be probed through either direct imaging of young planets \citep[e.g.][]{vanderburg2018detecting, ginski2018first}, or through photometric observations as they transit their host star \citep[e.g.][]{Heising15,Aizawa18,Kipping18}.
One such candidate is the young star 2MASS~J14074792-3945427 (`J1407') which underwent a complex eclipse two months in duration that was interpreted as the transit of a highly structured ring system filling the Hill sphere \citep{mamajek2012planetary, kenworthy2015modeling}.
%
%
In the case of planetary companions, transit photometry and spectroscopy of such a Hill sphere system provides the opportunity to probe both the spatial and chemical composition of a circumplanetary disk during planetary formation.

Alternatively circumstellar material can also periodically eclipse young stars, allowing us to probe stochastic processes in protoplanetary disks.
Many young stars have been observed to display such "dipper" behaviour \citep[][etc.]{Bouvier99,Cody14,Ansdell16a,Ansdell18}, and proposed explanations includes the transit of accretion streams \citep{Bouvier99}, material from asteroid collisions \citep{kennedy2017transiting}, coalescing circumstellar dust clumps \citep{rodriguez2013occultation}, etc.

%
%

PDS\,110 (HD~290380) is a young ($\sim11$\,Myr old) T-Tauri star in the Orion OB1 Association that showed two extended (2 week) eclipses (30\%) in 2008 and 2011, separated by a delay of 808 days.
An analysis by \citet{Osborn17} of all known photometry was consistent with an unseen companion in a periodic orbit of 808 days with a predicted 3-week long eclipse occurring around September 2017, although aperiodic UX Ori-like dimmings could not be ruled out.
If periodic, the  resulting ephemeris predicted two eclipses to have already occurred (in 2013 and 2015) however, due to the unfavourable placement of PDS~110 during this season, they were not observed by any photometric survey.
An observable eclipse was predicted at HJD=$2458015.5 \pm10$ ($1-\sigma$ region Sept 9-30 2017) with a full-width half maximum of $7\pm2$ days.

In Section \ref{sec:obs} we present photometry from a coordinated campaign\footnote{Co-ordinated at \href{http://pds110.hughosborn.co.uk}{http://pds110.hughosborn.co.uk}} to provide high cadence photometric measurements during the period from August 2017 into early 2018. \footnote{All photometry of PDS110 is available as supplementary material}.
%

In Section \ref{sec:spec} we detail further high-resolution spectroscopic observations obtained with TRES at the Whipple Observatory, and UVES on the VLT.
In Section \ref{sec:plate} we detail the analysis of nearly 40 years of photographic plates carrying out an archival search for other eclipse events.
With Section \ref{sec:discuss} and in the Conclusions we speculate what caused the observed eclipses and suggest future observations of PDS\,110.


\section{2017 Photometric Observations}
\label{sec:obs}
\begin{figure}
\centering
	\includegraphics[width=0.93\textwidth]{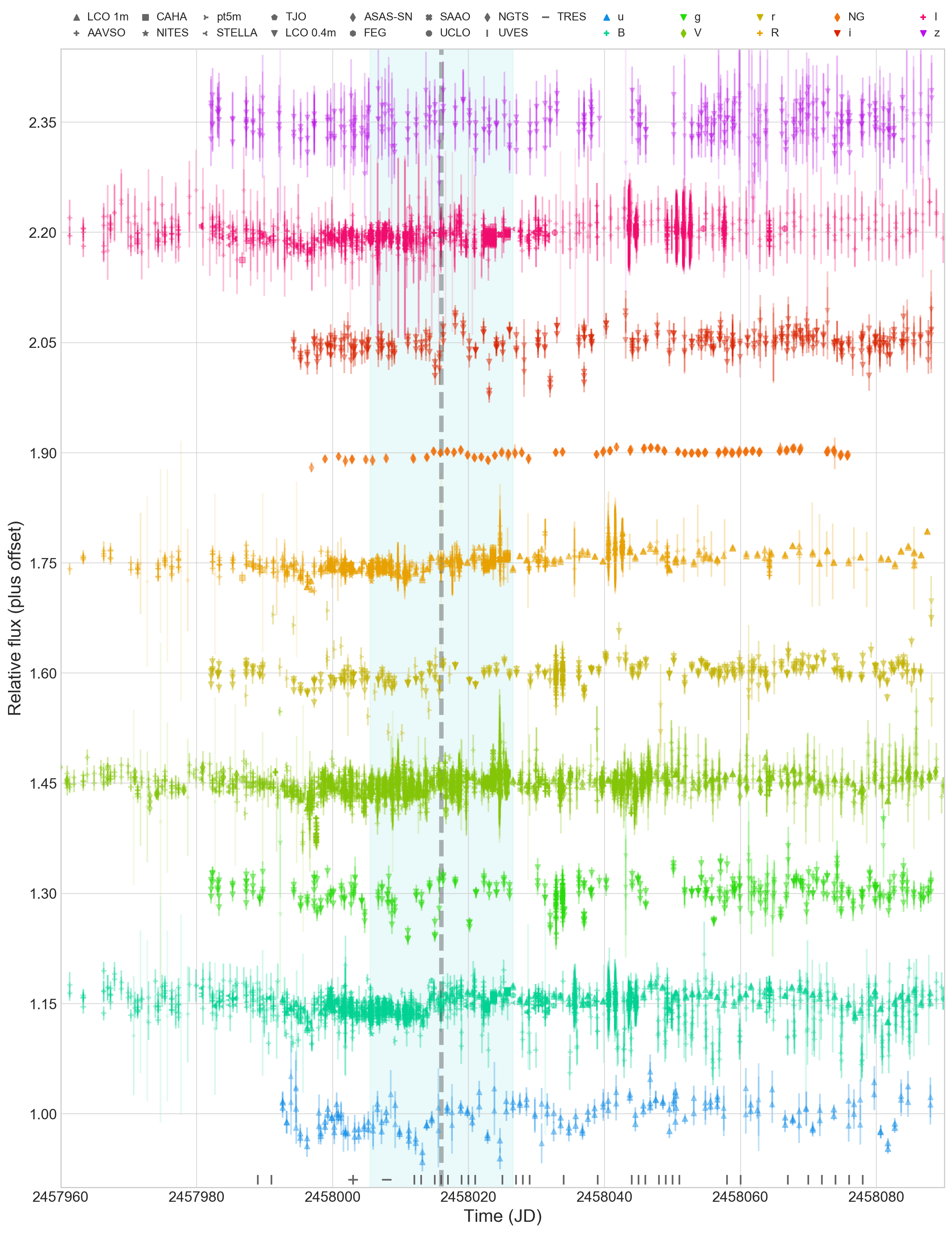}
    \caption{Photometry of PDS\,110 from JD=2457960 to 2458090, or 2017-Jul-25 to 2017-Dec-2. Telescopes used are shown using marker shape while filters are shown by colour and flux offset (with blue to red from bottom to top). $ugriz$ filters correspond to Sloan primed bandpasses. $UBVRI$ are Johnson/Cousins. Epochs of spectroscopic observations are shown at the base of the plot as vertical (VLT/UVES) and horizontal (TRES) lines. The transparency is dictated by the SNR, with points with large errorbars made fainter. The filled vertical region shows the predicted time of central eclipse from \citep{Osborn17} with the boundary corresponding to 1$\sigma$ uncertainties. Some observations continued into 2018 and are shown in Figure \ref{fig:dip2018}}
    \label{fig:big_figure}
\end{figure}

Photometric observations were taken by 11 professional observatories, with dozens more professional and amateur observers contributing through AAVSO.
These spanned 10 different optical filters including SDSS ugriz and Cousins BVRI filters, as well as the broad band NGTS filter.
The majority of observations began around 2457980 (2017 Aug 15) and finished once the time of predicted eclipse had past (2458090, or 2017 Dec 3).
These are summarized in Figure \ref{fig:big_figure}.
Some observations (from NGTS and AAVSO) continued into 2018, with a small part of that extended time frame shown in Figure \ref{fig:dip2018}.

In the following section we briefly summarize the observations of each contributing observatory.

\subsection{Contributing Observatories}
\subsubsection{Las Cumbres Observatory}
Las Cumbres Observatory (LCOGT) is a global network of robotic telescopes perfectly suited to the continuous, median-cadence observations required to detect long-duration dimmings of young stars.
Under the proposal ``Characterisation of the eclipsing body orbiting young star PDS 110'' (LCO2017AB-003), we were granted 35\,hr of time on the 0.4m network.
This consists of ten identical 0.4m Meade telescopes at six LCOGT observatory nodes: Siding Spring Observatory in Australia, Teide Observatory on Tenerife, McDonald Observatory in Texas, Cerro Tololo in Chile, SAAO in Sutherland, South Africa, and Haleakala Observatory in Hawai'i. 
These have a $2000 \times 3000$ SBIG STX6303 camera with a 14-position filter wheel including Sloan $u^\prime g^\prime r^\prime i^\prime z^\prime$, and Johnson/Cousins $V$ and $B$.
We primarily used 0.4m time to observe in Sloan $g^\prime r^\prime i^\prime z^\prime$. 

We were also assisted in these efforts by the observing campaign ``Time-Domain Observations of Young Stellar Objects (TOYS)'' (STA2017AB-002, PI: Aleks Scholz), which contributed 10\,hr of time on the 1m LCOGT network.
This includes telescopes at McDonald Observatory, Cerro Tololo, SAAO and Siding Spring Observatory.
These have a 4k $\times$ 4k Sinistro camera and 24 filter options including Johnson/Cousins $UBVRI$ and Sloan $u^\prime g^\prime r^\prime i^\prime$.
We primarily used the 1m time to observe PDS\,110 in Johnson/Cousins $BVRI$ and Sloan $u$ (where PDS\,110 is faintest).

In both 1m and 0.4m time, we took observing blocks of 3 images in each filter around 3 times per day, with exposure times adjusted to achieve SNR $\approx$ 200.
The data were accessed via the online observing portal, and the images and calibration files downloaded. 
AstroImageJ was then used to perform the calibrations and the reference photometry using the reference stars provided by AAVSO.

\subsubsection{AAVSO}
AAVSO is an international organisation designed to connect any observers capable of high-quality photometric observations (including amateurs) with astronomical projects which require observations \citep{kafka2016observations}.
An AAVSO Alert notice was released to observers \citep[alert 584, ][]{waagen2017}\footnote{\href{https://www.aavso.org/aavso-alert-notice-584}{https://www.aavso.org/aavso-alert-notice-584}}, which included a list of comparison stars, and more than 30 observers submitted observations during the campaign.

%
%

\subsubsection{NITES, La Palma}
The NITES (Near Infra-red Transiting ExoplanetS) telescope is a 0.4m, f/10 Meade telescope located at the Observatorio del Roque de los Muchachos on La Palma, and equipped with an e2v, $1024\times1024$ CCD with an FoV of $11.3\times11.3$ arcmin \citep{mccormac2014search}.
NITES observed PDS\,110 in four filters (Johnson $BVRI$) during 7 nights between JD=2457999 and JD=2458011. 
The \citet{mccormac2013donuts} ``DONUTS'' system enabled accurate auto-guiding.

\subsubsection{STELLA, Tenerife}
STELLA is composed of two 1.2m robotic telescopes at the Izana Observatory on Tenerife, Spain \citep{strassmeier2004stella} which focuses on long-term photometric and spectroscopic monitoring of stellar activity \citep[e.g.][]{mallonn2018gj1214}.
The wide-field imager, WiFSIP, has a $22\times22$ arcminute FoV and took observations of PDS\,110 four times per night in $B$, $V$ and $I$ filters (Johnson) with exposure times of 20, 12 and 10 seconds.
We obtained data on 38 nights from August to October 2017. The data reduction and extraction of the differential photometry of the target followed the description in \cite{mallonn2018gj1214}. We made use of \texttt{SExtractor} for aperture photometry and employed the same comparison stars for the three broad-band filters.

\subsubsection{NGTS, Chile}
NGTS (Next-Generation Transit Survey) is composed of 12x20cm telescopes, each observing 8.1 square degrees ($2.8^\circ \times 2.8^\circ$) of the sky with a wide-band filter (from 520 to 890nm) and a 2048x2048 deep-depleted CCD.
Its primary goal is to achieve mmag-precision photometry in order to search for transiting exoplanets \citep{wheatley2017next}.
Between Julian dates 2457997 and 2458199, PDS\,110 was included in one of the NGTS survey fields and continuously observed by a single camera while above 30$^\circ$ elevation. A of total of 95 nights of data were collected, with a typical hourly RMS below 1\%.
The raw 10 second NGTS frames were processed using a custom reduction pipeline \citep[][in prep]{choteinprep} to extract aperture photometry using several nearby comparison stars.
The data were binned to 1hr bins before being included with the other photometric data here.

\subsubsection{CAHA 1.23m, Calar Alto}
Remote observations enabled 251 images of PDS\,110 to be taken from the Calar Alto 1.23m telescope.
This robotic telescope has a DLR Mk3 CCD which observed in $BVRI$ Johnson filters.
Aperture photometry was performed with DEFOT \citep[see ][]{southworth09mn,southworth14mn} for PDS\,110 with three comparison stars providing relative photometry.

\subsubsection{ASAS-SN}
The All Sky Automated Survey for SuperNovae, or ASAS-SN \citep{shappee2014all,kochanek2017all} is a 20-unit network of wide-field telescopes designed to survey the entire sky in \textit{ugriz} \textit{g} magnitude down to magnitude 17 each night, with the primary goal of rapidly detecting supernovae.
We accessed ASAS-SN data of PDS\,110 data from the Sky Patrol search page\footnote{\href{https://asas-sn.osu.edu/}{https://asas-sn.osu.edu/}}.

\subsubsection{FEG, Sao Paulo}
Observations were carried out with a 16-in Meade LX200 telescope and a Merlin EM247 camera, with V-band filter and exposure time of 5 seconds. Useful data were acquired between 2017 September 2nd and 29th, totaling 5397 images in 14 nights. 

Each one of the $660\times498$ pixels frames was calibrated by bias subtraction and flat field correction. The fluxes of the target and nearby stars were determined from each image through aperture photometry taking advantage of the routines provided by the IDL Astronomy Library. The magnitude was calculated using the comparison stars provided by AAVSO (usually 000-BMH-803), with an error of 0.01 mag.

To determine the time evolution of the magnitude, the data were averaged every 36 images (3 minutes cadence), avoiding any spurious variation due to instrumental or meteorological effects.

\subsubsection{TJO, Montsec Astronomical Observatory}
PDS\,110 was observed with the Joan Or\'o robotic 0.8 m telescope (TJO) at the Montsec Astronomical Observatory in Catalonia. The TJO is equipped with Johnson/Cousins UBVRI filters and an e2v $2k\times2k$ CCD with a FoV of $12.3\times12.3$ arcmin. Johnson $B$ and $I$ filters were used and several observing blocks per night with 5 exposures each were configured. The exposure times for each filter were adjusted in order to achieve SNR $\approx$ 300. The images were reduced using the ICAT reduction pipeline at the TJO \citep{Colome2006} and differential photometry was extracted using AstroImageJ. The final TJO dataset contains 255 and 225 data points in the $B$ and $I$ filters, respectively, taken in 20 different nights between September 5th and October 9th.

\subsubsection{pt5m, La Palma}
\textit{pt5m} is a 0.5m robotic telescope located at the Roque de los Muchachos Observatory, La Palma \citep{hardy2015pt5m}. It observed PDS\,110 on 21 separate nights between JD=2457993 and 2458015 in Johnson $B$, $V$ and $R$ filters. Astrometry was performed automatically on all images by cross matching detected sources against the 2MASS point-source catalog. Instrumental magnitudes were calculated for all detected objects in the images using {\sc sextractor}. Instrumental magnitudes for the $B$ and $V$ observations were calculated using zero-points derived by cross-matching against the APASS catalogue, whilst a cross-match against catalogued SDSS-$r^\prime$ magnitudes gave a zeropoint for the $R$-band images. No colour terms were applied.

\subsubsection{SAAO}
The SAAO 1m was used on two nights to observe PDS\,110 in 3 bands using a Sutherland high-speed optical camera \citep{coppejans2013characterizing}.
However the small field of view ($2.85 \times 2.85$ arcminutes) made reference stars difficult, and the reduction required the use of measurements submitted by other observatories for calibration.
The high-cadence data (cadence from 0.7 to 10s) allowed a search for short-period oscillations ($P<3d^{-1}$), however none were detected.
The data were binned with a weighted mean to 7.2-min bins before being included in the ensemble analysis.

\subsubsection{UCL Observatory}
PDS\,110 was observed on eleven separate nights between JD 2457996 and 2458165 from the University College London Observatory (UCLO), located in Mill Hill, London. A fully robotic 0.35-m Schmidt Cassegrain was used with a SBIG STL-6303E CCD camera. Observations were taken in Astrodon $R_c$ and $I_c$ (Cousins) filters \citep[for more observing details see][]{fossey2009detection}. Typically, 10--30 exposures of 20 seconds were obtained in each filter on each night; differential photometry relative to an ensemble of nearby comparison stars yielded a total of 230 measurements in $R_c$ and 150 in $I_c$, binned to provide average relative fluxes on nine nights for each filter.

\subsection{Photometric ensemble analysis}
With any observing campaign involving the inclusion of photometry between multiple observatories across multiple filters, the pooling and comparison of data is a difficult task.
Each observer introduces their own systematics, including most visibly an offset in the magnitude or normalised flux level.
This is despite, in some cases, using identical filters and the same comparison stars\footnote{\href{https://www.aavso.org/apps/vsp/chart/X20500AB.png}{Provided by AAVSO}}.
In the case of our PDS\,110 campaign, however, the precise magnitude measurements are not as important as the relative change over time.
Therefore we applied an offset to each lightcurve to enable comparisons between them, using the long-baseline and high accuracy of the LCOGT photometry as a guide.
In the case where lightcurves were provided with normalised flux, we converted these to differential magnitudes taking the archival magnitude as the whole-lightcurve flux median before assessing the offsets.

The potential low-level variability of PDS\,110 and the large variations in observation cadence between observations mean that simply adjusting the medians of data in a certain region is not ideal.
Instead, we developed a minimisation process which computes the sum of the magnitude difference between each point on one light curve and each point on another ($y_{a,i}-y_{b,j}$ in Eq 1 where y is magnitude and $a$ and $b$ represent two photometry sources).
This is then weighted for the time separation between those points ($x_{a,i}-x_{b,j}$ in Eq 1 where x is time in days).
In an effort to remove the influence of a structured lightcurve combined with irregular time-sampling, we weighted the magnitude difference between points by the absolute time difference between them, scaled using a squared exponential and a lengthscale ($l$) of 4 days.
The minimisation function ($f_{\rm min}$) is defined in Equation \ref{eq:1}.
\begin{equation}
f_{\rm min} = \sum_{i=1}^{N_a} \sum_{j=1}^{N_b} \frac{\left(y_{a,i}-(y_{b,j}+\Delta_m)\right)^2}{\sigma_{a,i}^2+\sigma_{b,j}^2} \exp{\frac{- (x_{a,i}-x_{b,j})^2)}{2 l^2}}
\end{equation}\label{eq:1}
Bootstrapping was performed to assess the increase in errors due to this method, which were added in quadrature to the flux of the adjusted points.
This procedure was then performed iteratively on each dataset in each filter until the offsets converged, with the exception of our LCOGT data (and CAHA data in $I$-band), which we held as a fixed reference lightcurve. 
The result is a magnitude offset ($\Delta m$) and uncertainty for each filter, and for each telescope.
NGTS data were not minimised in this way as it observed in a unique broadband filter.


The computed offsets for each telescope, which have been converted to relative flux to match the lightcurves presented in the following figures, are shown in Table~\ref{tab:photometry}. 
They show good agreement for the $B$- and $V$-band, but large negative shifts in relative flux for $R$ and $I$, suggesting a disagreement between the historic $R$- \citep{2003yCat.1289....0Z} and $I$-band values \citep{2005yCat....102002D} which the baseline LCOGT data were adjusted to.
However, as we are focused on the change in time, these variations are unlikely to cause significantly increased systematics.

Full photometry for PDS\,110 during the campaign is shown in Figure 1.
We also release all data publicly as supplementary material to this publication.

\onecolumn

\twocolumn

\begin{figure}
	\includegraphics[width=\columnwidth]{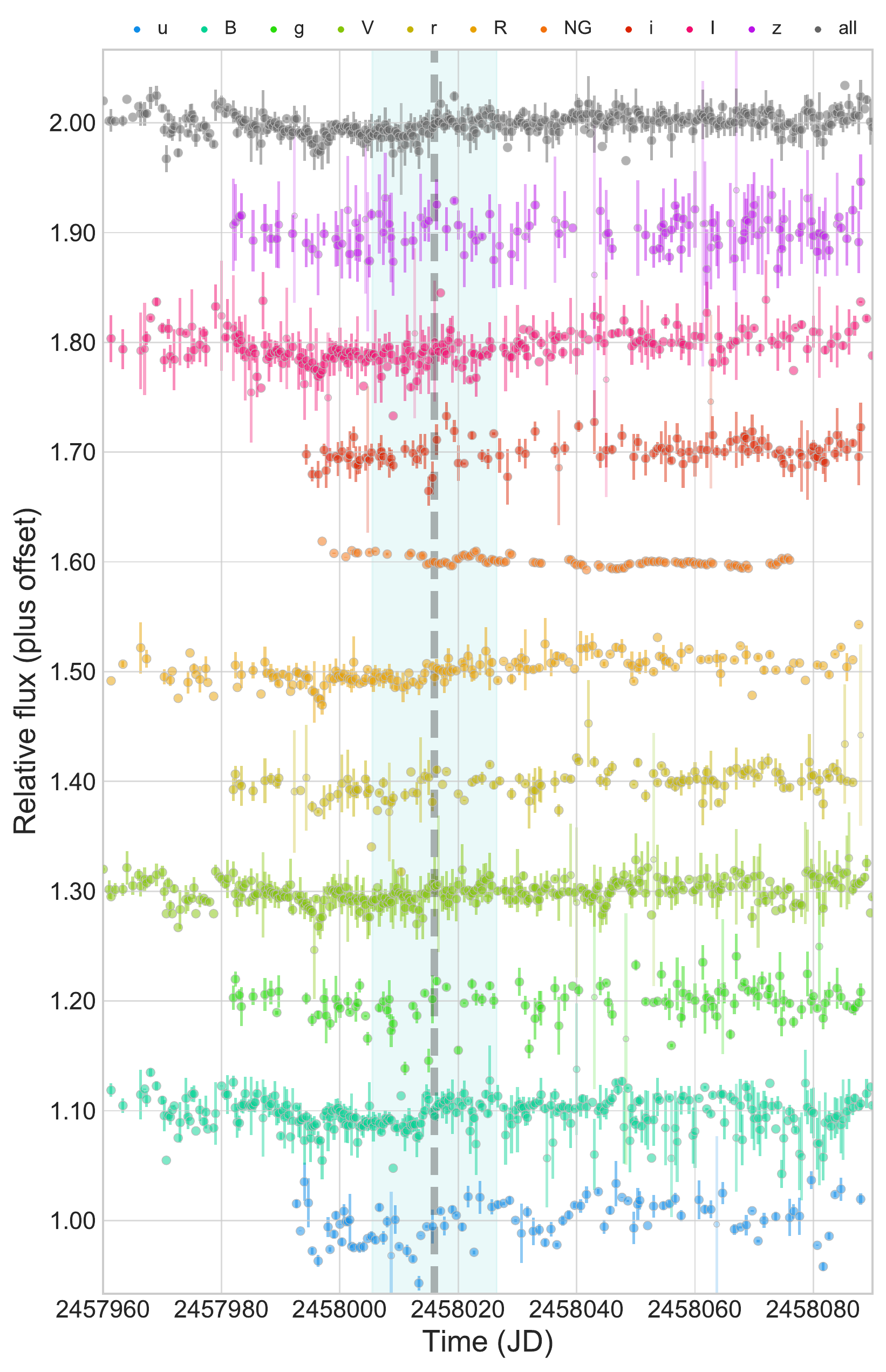}
    \caption{Photometry of PDS\,110 binned into 0.333d time bins for each filter and spanning the same time period as Figure \ref{fig:big_figure}. The combined lightcurve for all filters is shown above in grey.}
    \label{fig:binned_figure}
\end{figure}

\subsection{Observed candidate dimming events}
Two significant dimming events were observed, although their occurrences are inconsistent with the prediction from previous dimmings, in terms of both timing and depth.
The first was before the predicted time of eclipse at JD$\sim 2457996$ in all bands (visible in the binned photometry in Figure \ref{fig:binned_figure}).
It lasted less than 1 day and saw flux dip by only $\sim5\%$, so does not resemble the previously reported events.

\begin{figure}
	\includegraphics[width=\columnwidth]{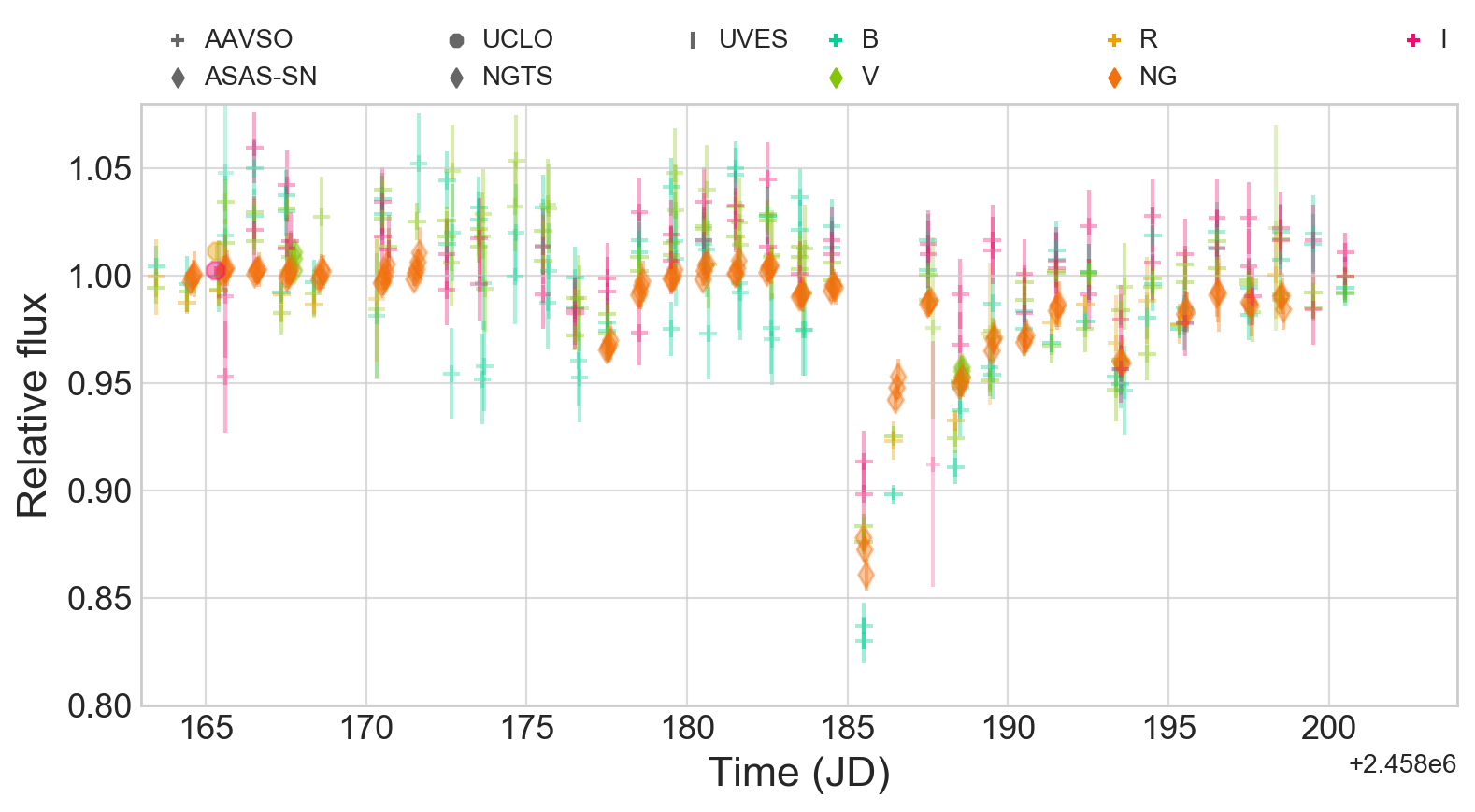}
    \caption{Photometry (with no flux offset) of the short-duration eclipse seen in 2018.}
    \label{fig:dip2018}
\end{figure}

A second dimming event was seen after the official end of the campaign in 2018 with a centre at $\rm{JD}=2458186$ (see Figure \ref{fig:dip2018}).
Similarly, its shape is for the most part inconsistent with the previously observed dimmings - it is both far weaker and of shorter duration, with only a single night showing a depth,$\delta>10\%$.
While the NGTS data show the event clearest, it was also observed by AAVSO observers and ASAS-SN.
These also show that shallower dips (of $\sim 4\%$) occurred $\sim8$ days before and afterwards. 

These two events appear to suggest that more rapid-timescale dimmings are possible than expected from \citet{Osborn17}, and may suggest the single-night flux drops observed in ASAS data in 2006 and 2007 may have also been real rather than, as speculated in \citep{Osborn17}, anomalous flux values.

\subsection{Reddening}
\begin{figure}
	\includegraphics[width=\columnwidth]{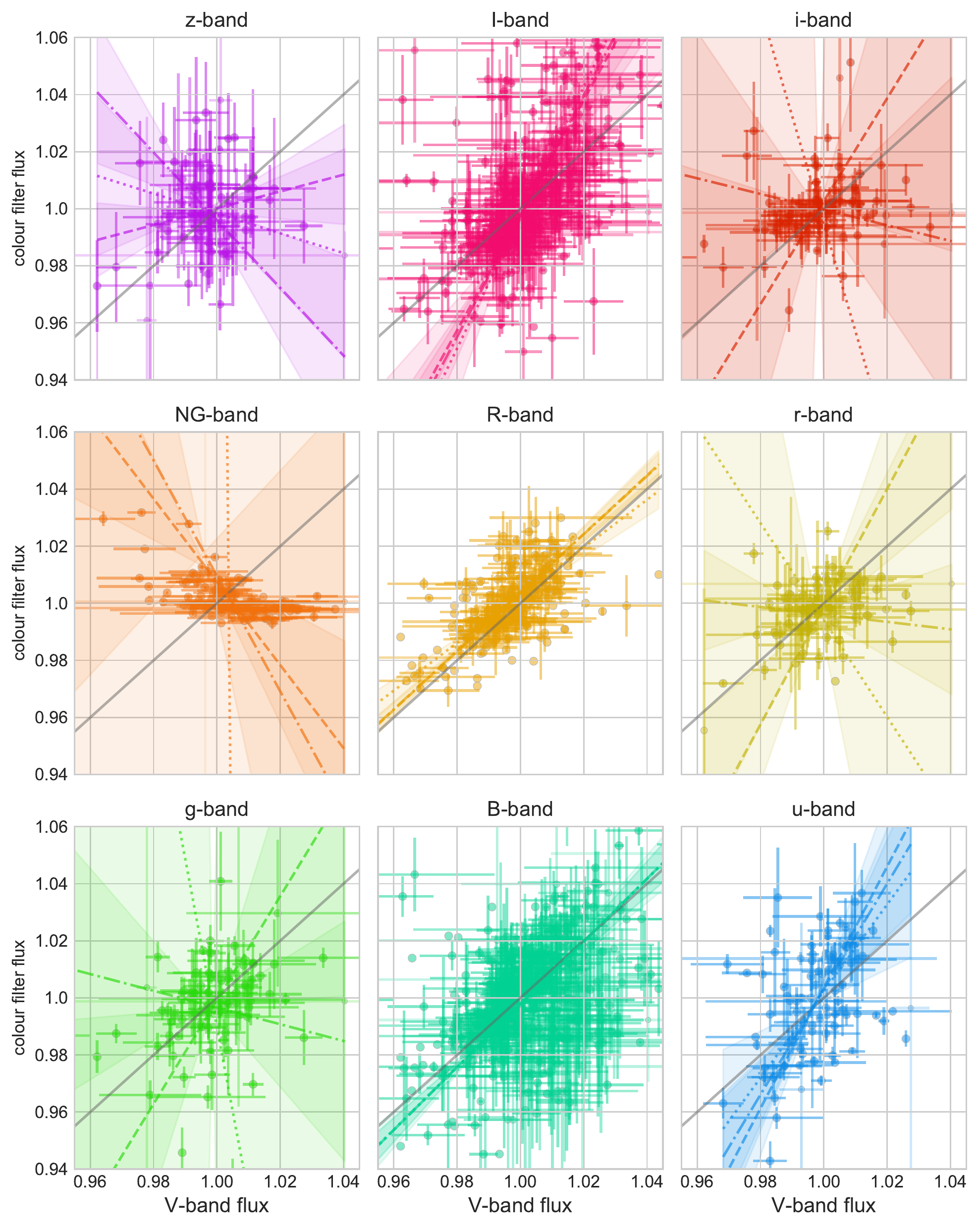}
    \caption{Flux correlations for each binned filter values compared to V-band. Points along the diagonal line (shown in grey) would represent dimmings perfectly correlated with V-band flux (and therefore "grey" ). Lines of best fit are computed using \texttt{bces} and "Orthogonal least squares" (dashed), Vmag as the independent variable (dotted), and the  bissector method (dash/dotted). $1-\sigma$ error regions for each are overplotted.}
    \label{fig:correl_figure}
\end{figure}

Obscuration of the star by small dust causes more light to be blocked by dust grains close in size to the wavelength of light.
Hence, typically, dips appear deeper in blue filters than in red. 
Although no major dips were observed, short duration and shallow depth variability seen in PDS\,110 may be enough to spot the imprint of dust.
In Figure \ref{fig:correl_figure} we explore this by plotting the difference in magnitude of the binned $V$-band lightcurve (our most well-covered filter) and the photometry from other filters taken at the same time.
Lines of best fit are plotted using the \texttt{bces} package\footnote{\href{https://github.com/rsnemmen/BCES}{https://github.com/rsnemmen/BCES}} and three assumptions detailed in the figure caption \citep{2012Nemmen}.
We see that the gradient in the $u$-band appears far steeper than would be expect for a ``grey'' absorber.
Intriguingly, the I-band observations also show a steeper-than-gray correlation, especially due to brighter-than-average points. This remains unexplained and appears to contradict the effect of reddening. Systematics, especially for the low-SNR photometric observations in the I-band, would appear the most likely cause.

However, correlations are also likely present due to telescope-specific systematics across all bands and times, which would similarly manifest as a positive correlation between filters. 
This may be responsible for why $BRI$ filers show stronger correlations to $V$-band than $ugriz$ filters (which were typically not observed contemporaneously as $V$).
Therefore we choose not to model the reddening present in all observations, although we note that dust may be present.
Exploring the extinction or dust grain analysis of single dips (eg that in Figure \ref{fig:dip2018}) is also problematic due to the lack of perfectly simultaneous data and uncorrected systematic offsets between telescopes.

\section{High resolution spectroscopy}\label{sec:spec}
\subsection{TRES}
Using the Tillinghast Reflector Echelle Spectrograph \citep[TRES;][]{furesz:2008}\footnote{\url{http://www.sao.arizona.edu/html/FLWO/60/TRES/GABORthesis.pdf}} on the 1.5 m telescope at the Fred Lawrence Whipple Observatory on Mount Hopkins, AZ, we observed PDS\,110 nine times from UT 2016 Oct 09 until UT 2017 Sep 11. The spectra were taken with a resolving power of $\lambda / \Delta \lambda \equiv R = 44000$ covering a wavelength range of $3900-9100$\AA. 
For each order, we cross-correlate each spectrum against a template made from all median-stacked spectra that is aligned to that with the highest S/N To derive the relative RVs, we fit the peak of the cross-correlation function across all orders. The scatter between each order for each spectrum determines the uncertainties on the relative RVs \citep{Buchhave:2010}. Activity and rotation mean that the resulting relative RVs give a uniform offset from that initial high-S/N spectra, therefore we re-adjust the RVs to be self-consistent. We also performed a fit simply using the strongest observed spectra as a template, which gives consistent results but with slightly lower precision.

We see no large variation ($>$1 km s$^{-1}$) in the TRES radial velocity measurements. We note that since PDS\,110 is a late-F star with broad lines due to a projected equatorial rotational velocity of 60 \kms, precise radial velocities are challenging. 
Our observations cover the first half of the predicted orbital period from \citet{Osborn17} with a standard deviation from the mean of 0.53 \kms. 
Using a 3$\sigma$ value as the upper limit (1.59 \kms), assuming a 1.6\msun\ host star, and fixing the orbit to that of the predicted ephemeris (T$_C$ = 2454781, P= 808.0 days) from \citet{Osborn17}, this would correspond to an upper mass limit for the companion of $\sim$100 M$_{\rm Jup}$. We also run a Levenberg--Marquardt fit to the RVs, enforcing the ephemeris and a circular orbit, and get a 3$\sigma$ upper limit on the mass of 68 M$_{\rm Jup}$. However, these upper limits make the assumption that we know the ephemeris of the companion. The RV measurements from TRES are shown in Table \ref{tab:tresrvs}. 


\begin{table}
\footnotesize
 \centering
 \setlength\tabcolsep{7pt}
\caption{TRES Relative Radial Velocity Measurements}
\begin{tabular}{c c c}
 \hline
\bjdtdb &   RV & $\sigma$RV   \\
          & (\kms) & (\kms)  \\
 \hline
2457670.98905&       -0.32 &     0.35 \\
2457679.98738&       -0.98 &     0.45\\
2457685.00057&       -1.02 &     0.37\\
2457786.70741&        0.00 &     0.38\\
2457800.69517&        0.15 &     0.43\\
2457823.72000&       -1.22 &     0.42\\
2457855.63204&       -0.46 &     0.33\\
2458002.98522&       -0.90 &     0.58\\
2458007.98572&        0.10 &     0.27\\
\hline
\end{tabular}
 \label{tab:tresrvs}
\begin{flushleft}

\end{flushleft}
\end{table}
\subsection{UVES}
High spectral resolution observations of PDS\,110 were obtained with the Ultraviolet and Visible Echelle Spectrograph~\citep[UVES;][]{2000SPIE.4008..534D} as part of the DDT programme 299.C-5047 (PI: De Mooij) on 32 nights between August 24 and November 21, 2017.

By using the \#2 Dichroic, the spectra on every night were obtained using both the blue and red arms simultaneously with the 437+760nm wavelength-setting. Using this setup, the blue arm covers a wavelength range from $\sim$3730\,\AA\ to $\sim$5000\,\AA, while the red arm has a wavelength coverage from $\sim$5650\,\AA\ to $\sim$9560\,\AA, with a small gap between the two CCDs that make up the red array. In this paper, however, for the red arm we only use the shorter wavelength CCD, as this is less affected by telluric lines. The blue arm is not affected by telluric lines.
During each visit, a total of four spectra were obtained, each with an exposure time of 300 seconds.

The data were reduced using the ESO UVES pipeline version 5.7.0 through ESO Reflex. The pipeline reduced and merged spectra from each epoch were combined to increase the signal-to-noise ratio.
As the wavelength range of the red arm of UVES contains strong telluric bands (including the O$_{2}$ bands), we first used the ESO Molecfit tool\footnote{https://www.eso.org/sci/software/pipelines/skytools/molecfit} \citep[][]{SmetteEtAl15} to correct the spectra for telluric absorption.

\begin{figure}
\includegraphics[width=\linewidth]{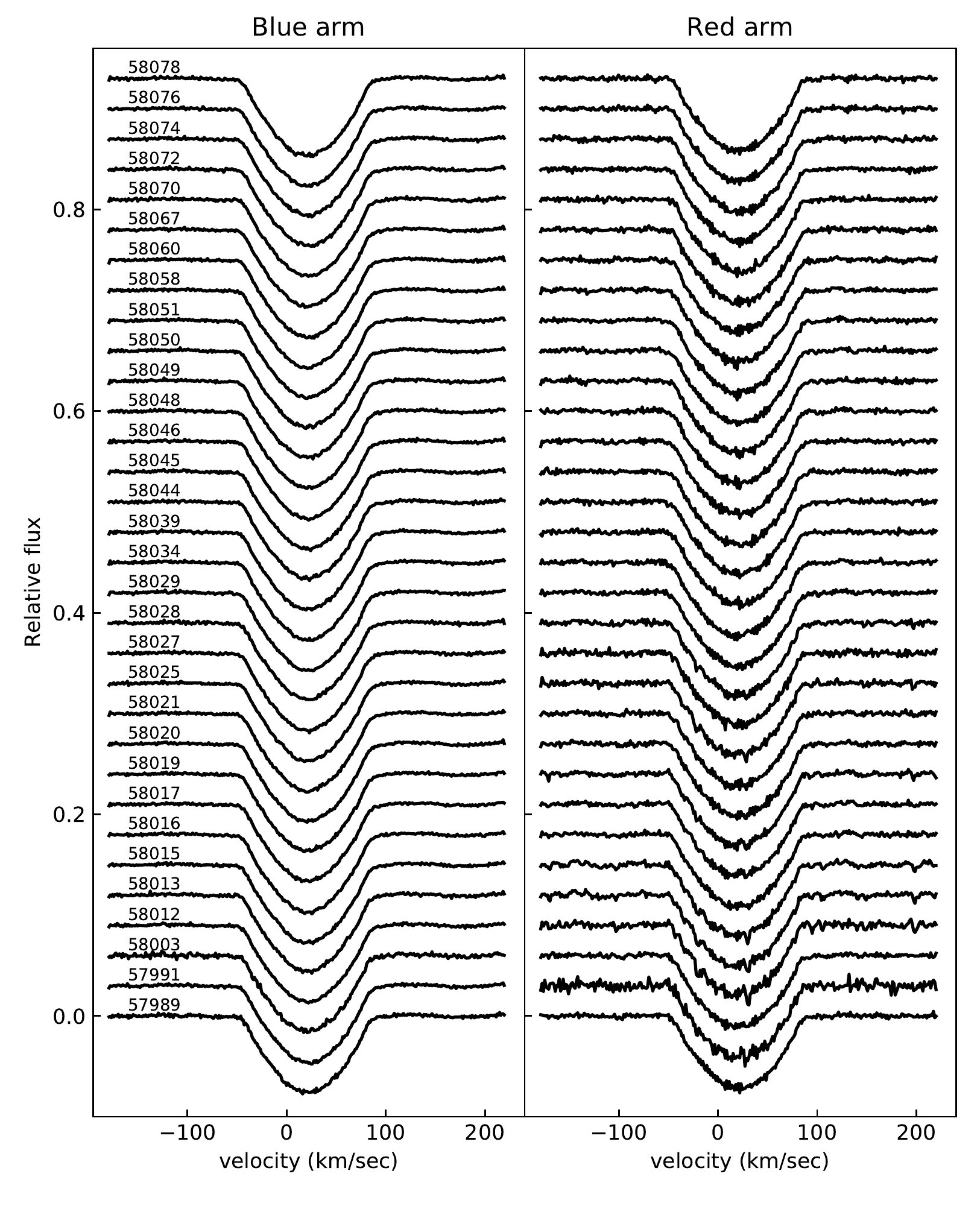}
\caption{\label{fig:UVES_lprof}Least-Squares Deconvolution line-profiles of the UVES observations of PDS 110. The Julian dates for the start of the night are indicated above each profile. The left panel shows the profiles for the blue arm of UVES, while the right panel shows the same for the red arm. }
\end{figure}

\begin{figure}
\includegraphics[width=\linewidth]{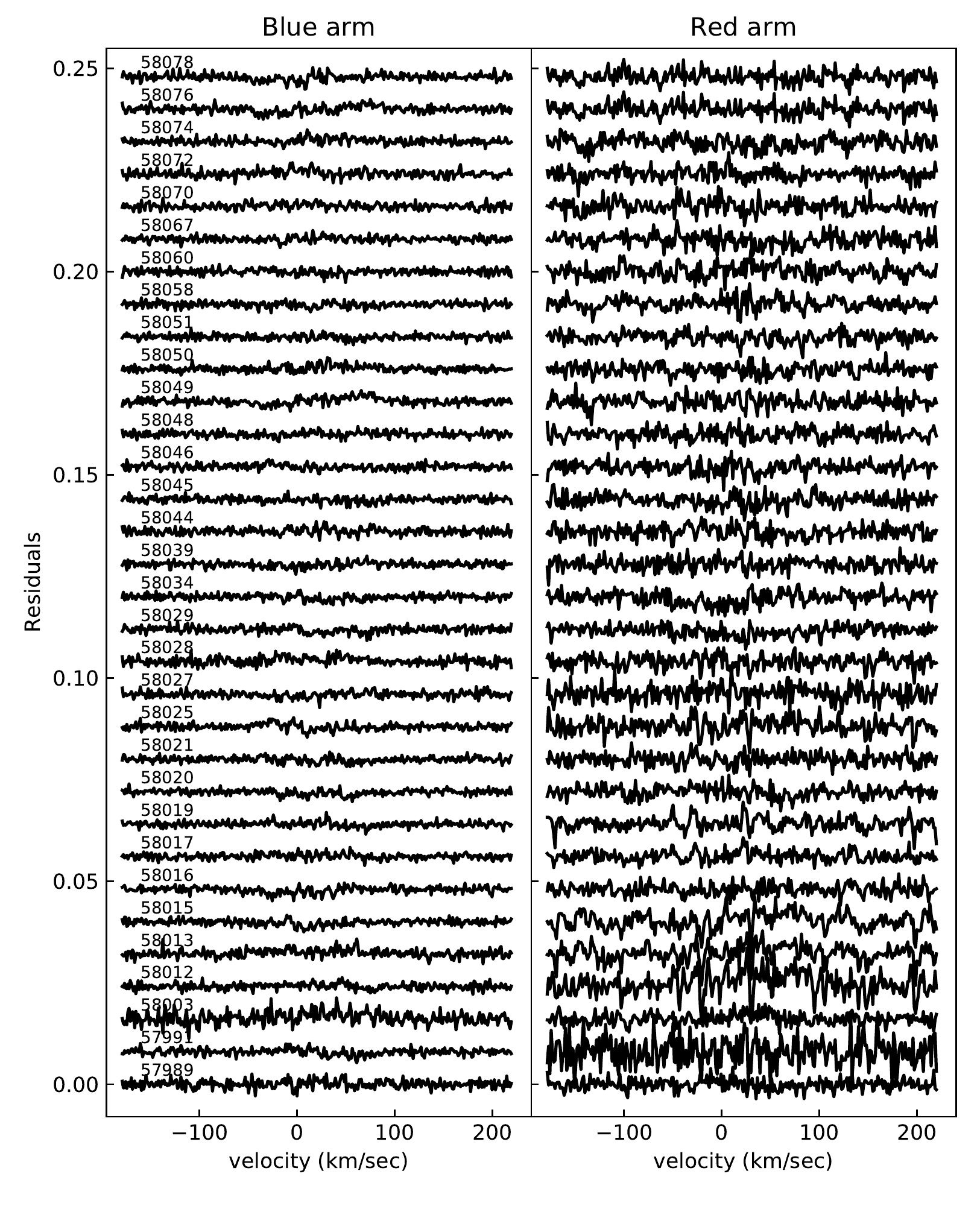}
\caption{\label{fig:UVES_resid}Differences of the line-profiles shown in Fig.~\ref{fig:UVES_lprof} compared to the median line-profile over the entire UVES observing campaign.}
\end{figure}

The observations were corrected for blaze-variations from epoch to epoch, by first dividing the spectra from each epoch by the spectrum of the first epoch, binning this ratio, interpolating it using a cubic spline, and finally dividing the spectra by the interpolated function. This was done for each of the arms separately. A master spectrum was generated by averaging the blaze-corrected spectra from individual nights, and the envelope was used to create a continuum normalisation that was then applied to all spectra. Finally, we used Least-Squares Deconvolution, based on \cite{DonatiEtAl97} as implemented by \cite{WatsonEtAl06}, to combine the signal from the multiple stellar lines and increase the signal-to-noise ratio. Care was taken to mask both bands with strong telluric residuals (e.g. the saturated O$_2$ bands in the red arm) as well as wavelength regions that are strongly affected by stellar emission features (e.g. the Balmer lines, Ca II H\&K lines, the Na D lines) due to accretion. The linelist of $\sim2400$ lines was generated using the 'Extract Stellar' option from the VALD3 database \citep{VALD3}\footnote{http://vald.astro.uu.se/} for the stellar parameters from \cite{Osborn17}. 
The resulting LSD profiles for the red and blue arms are shown in Fig.~\ref{fig:UVES_lprof}. In Fig.~\ref{fig:UVES_resid} we show the differences between the individual line-profiles and the median line-profile taken over the entire UVES observing campaign.
Structure transiting the stellar disk (e.g. a ring-crossing event) would induce a bump in the (residual) line-profile where light from the stellar surface at a certain Doppler shift is occulted, \citep[which causes the Rossiter-McLaughlin effect, e.g.][]{DeMooijEtAl17}, however, no such signature is observed. 
A detailed study of the emission lines, which show information about the accretion rate, the inclination of the star, etc, will be included in a future analysis \citep{DeMooijEtAlPrep}.


\begin{figure}
\includegraphics[width=\linewidth]{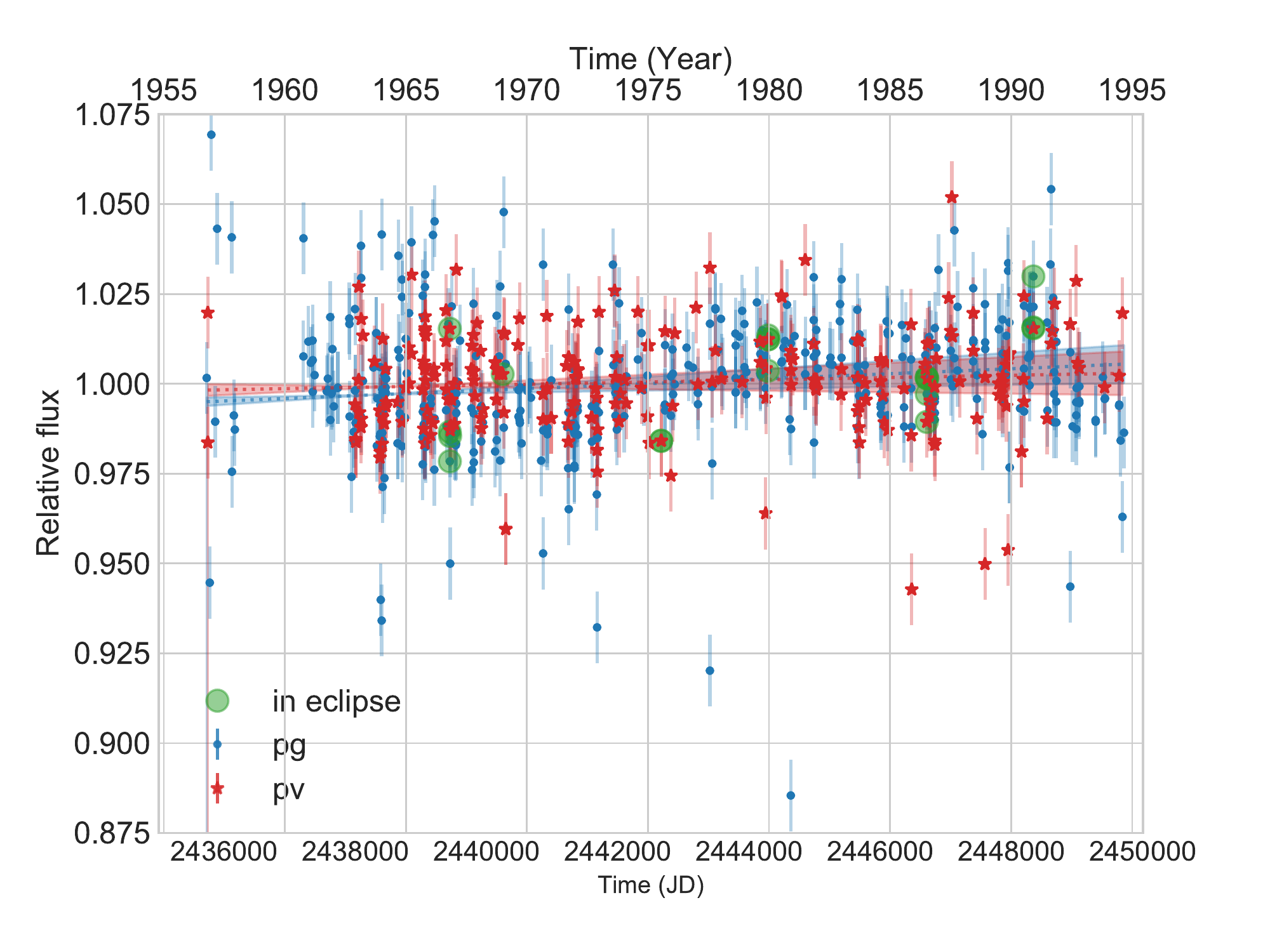}
\caption{\label{figure_sonneberg_timeseries}Sonneberg plate archival photometry from 1956 to 1994 in  \textit{pg} (blue) and \textit{pv} (red) filters. We use the point-to-point median absolute difference ($\sim1\%$) as a global uncertainty, as individual measurement uncertainties are typically underestimated and likely systematics-dominated. Points circled represent those predicted to be in-eclipse using the ephemeris of \citep{Osborn17}. Dashed orange and blue lines show 1D polynomial trends, and filled regions show the $1\sigma$ error cones for each}.
%
%
\end{figure}

\section{Plate photometry 1956-1994}
\label{sec:plate}
The second largest plate archive in the world, after Harvard (which has yet to digitize data from PDS\_110), is located at Sonneberg Observatory \citep{1992Stern..68...19B}. Two observation programs contributed $275{,}000$ plates between 1935 and 2010 in two colour bands, \textit{pg} (blue) and \textit{pv} (red) \citep{1999AcHA....6...70B}. We use the Sky Patrol plates of $13\times13$~cm$^2$ size, a scale of 830 arcsec per mm, giving a field size of about $26^{\circ}\times 26^{\circ}$, taken between 1935 and 1994. The limiting magnitudes are of order 14.5\,mag in \textit{pg} and 13.5\,mag in \textit{pv}. Plates were scanned at 15 $\mu$m with 16 bit data depth. Typical exposure times are 30 to 60 minutes.

Our reduction pipeline is described in-depth in \citet{2017ApJ...837...85H}. In brief, we perform an astrometric solution \citep{2010AJ....139.1782L} using a list of coordinates of the brightest sources as an input and the Tycho-2 catalog as a reference. With the source coordinates, we perform photometry using the SExtractor program \citep{1996A&AS..117..393B} with a constant circular aperture.

As quality filters we remove plates with suboptimal astrometric solutions, and those with bad quality after visual examination, which included all plates between 1936 and 1956, potentially to plate degradation. For calibration, we used the ten nearest stars between magnitude 10 and 12, as recommended by the AAVSO observation campaign. After calibration, the average standard deviation of the magnitudes is $\sim$0.05\,mag, significantly better than the $\sim$0.1\,mag obtained on plates for dimmer stars \citep[e.g.,][]{2009AJ....138.1846C, 2010PZ.....30....4G,2014ApJ...780L..25J}. We attribute the better quality to stricter quality cuts, the higher brightness of the star, and its location near the plate centre on many plates.

We show the time-series photometry in Figure~\ref{figure_sonneberg_timeseries}, where blue and red symbols represent the photometric bands. Our good data covers JDs 2435730 to 2449710, or dates between September 14, 1956 and Dec 24, 1994. There are no obvious dimmings in the timeseries, with the darkest measurement $\sim10\,$\% below the mean. No significant long-term trend is detected in either filter (trends are $7.5\pm2.3 \times 10^{-9}yr^{-1}$ and $3.3\pm2.8\times 10^{-9}yr^{-1}$ in pg and pv respectively), suggesting the brightness of PDS~110 is stable on the order of decades. 

When we phase-fold the data to a period of 808 days, a handful of data points are near (within $15d$ of) the expected eclipse, as shown in Figure \ref{figure_sonneberg_timeseries}. In no case do we see any indication of a dip.

\section{Discussion}
\label{sec:discuss}
The ephemeris predicted in \citet{Osborn17} relied on the detection of two bona fide dips, plus a lack of corroborating photometry at other predicted eclipse times.
However the photometry collected by our campaign reveal no dip with a depth greater than $\sim1\%$ during the predicted ephemeris (HJD=$2458015.5$ or 2017-09-20 $\pm10$days).
One potential solution to the lack of an event may be that the orbit of the body has decayed such that the eclipse was missed. 
However, extensive pre-dip data in 2013 (eg with KELT and ASAS-SN photometry), the archive photometry from Sonneberg plates, and the long-baseline of the 2017 observations helps rule out this hypothesis.

Such a rapid movement of large dust structures on the timescale of only a few orbits would contradict the hopeful hypothesis of \citet{Osborn17}, which postulated long-lived dust encircling a periodic giant planetary or low-mass stellar object.
The absence of an RV signature (albeit in noisy, rotation-dominated data) also points away from any hypothesis involving a high-mass companion.
In sum, we no longer have substantial proof of PDS\,110's periodicity and the data are more consistent with an aperiodic explanation.
  
The presence of other smaller (and shorter-duration) dips, two of which were observed during the 2017-2018 observing campaign (see Figures \ref{fig:binned_figure} \& \ref{fig:dip2018}), and some of which were hinted at in ASAS 2006 observations, also suggest an aperiodic cause.
The low-flux points seen in archival Sonneburg photometry (see Figure \ref{figure_sonneberg_timeseries}) may also be the result of bona fide short-duration dipping events, unresolved due to the $\sim$few day cadence of those observations.

PDS\,110 is encircled by a large dust disk, as revealed in the IR observations, and this dust is likely the source of any deep and short-duration variability.
The lack of reddening suggests we are observing PDS\,110 high above the disk plane, and therefore some mechanism must exist to get clumps of material into our line-of-sight, some large enough to block 30\% of the starlight for days, as in 2008 and 2011.
The exact structure of the dust disk could be revealed using high-resolution sub-mm imaging \citep[e.g. with ALMA, as was performed for dipper star EPIC 204278916,][]{scaringi2016peculiar}.

Large-scale version of these aperiodic dimmings have been observed as UX Ori type variables, such as the dips of AA Tau \citep{bouvier2003eclipses}, V1247 Orionis \citep{caballero2010occultation}, RZ Psc \citep{kennedy2017transiting}, and V409 Tau \citep{Rodriguez2015}.
Similar dips with an unexplained origin have also been seen around older stars, for example KIC 8462852 \citep{boyajian2016planet}.

The quantity of photometry assembled for PDS\,110 here and in \citet{Osborn17} reveal that dimming events are exceedingly rare, with dips greater than a few percent in depth occurring during at most 2\% of the time.
The events are also typically far shallower in magnitude than a typical UX Ori. 
Therefore, maybe we are seeing such a system at an extremely high viewing angle, at extremely low optical depth, or potentially at a dissipative stage of UX Ori evolution.
A more detailed exploration of the high-resolution spectra obtained by UVES and TRES during this campaign may help answer the question of what caused the aperiodic dips of PDS~110 \citep{DeMooijEtAlPrep}.
Alternatively, the increasing quality of ground-based \citep[e.g.][etc.]{wheatley2017next, shappee2014all} and space-based \citep{ricker2010transiting} photometry may reveal more low-amplitude UX Ori systems like PDS\,110.






\section{Conclusions}

A large ground-based follow-up campaign of PDS\,110 was conducted to search for the predicted eclipse of a dust-encircled massive body postulated to be orbiting within (or above) the dust disk of PDS\,110.
This included a dozen professional observatories and more than 30 amateur observers. 
The high-quality photometry recorded spans 10 filters and more than 200 days. 

This campaign, and the lack of any eclipse at the predicted transit time, has allowed us to rule out the hypothesis that PDS\,110 has a dust-enshrouded companion.
This is also backed up by archival photometry from Sonneburg archive, which does not reveal dimmings at the predicted times, radial velocity observations from TRES, which sees no signal from a stellar companion, and UVES observations of PDS\,110 during the predicted eclipse, which see no variation in the stellar line profiles.
However, the photometric campaign did reveal that PDS\,110 does undergo shorter and/or shallower dimming events.

Together, the observations point to a new, aperiodic source of the eclipses, potentially from dust blown above the disk-plane as has been hypothesised for UX Ori-type variables. 
Future observations of PDS\,110 may reveal more such events, and future all-sky surveys may detect more PDS\,110-like eclipsers.

\section*{Acknowledgements}
This research made use of Astropy,\footnote{http://www.astropy.org } a community-developed core Python package for Astronomy \citep{astropy:2013, astropy:2018} and of Matplotlib\citep{Hunter:2007}.
We thank our anonymous referee for their helpful comments.
We acknowledge with thanks the variable star observations from the AAVSO International Database contributed by observers worldwide and used in this research.
HPO acknowledges support from Centre National d'Etudes Spatiales (CNES) grant 131425-PLATO.
M.H. thanks Frank (Theo) Matthai for assistance with finding the relevant Sonneberg plates in the observatory archive. This work has made use of the VALD database, operated at Uppsala University, the Institute of Astronomy RAS in Moscow, and the University of Vienna. 
E.H. acknowledges support by the Spanish Ministry of Economy and Competitiveness (MINECO) and the Fondo Europeo de Desarrollo Regional (FEDER) through grant ESP2016-80435-C2-1-R, as well as the support of the Generalitat de Catalunya/CERCA programme. 
The Joan Or\'o Telescope (TJO) of the Montsec Astronomical Observatory (OAdM) is owned by the Generalitat de Catalunya and operated by the Institute for Space Studies of Catalonia (IEEC). 
pt5m is a collaborative effort between the Universities of Durham and Sheffield. The telescope is kindly hosted by the Isaac Newton Group of Telescopes, La Palma.
GMK is supported by the Royal Society as a Royal Society University Research Fellow. 
TB and RWW acknowledge support from the UK Science and Technology Facilities Council (STFC; reference ST/P000541/1).
PC acknowledges funding from the European Research Council under the European Union's Seventh Framework Programme (FP/2007-2013) / ERC Grant Agreement n. 320964 (WDTracer).
This paper uses observations made at the South African Astronomical Observatory (SAAO).
ASAS-SN light curves are primarily funded by Gordon \& Betty Moore Foundation under grant GBMF5490.
M.N.G. is supported by the STFC award reference 1490409 as well as the Isaac Newton Studentship.
L.M. acknowledges support from the Italian Minister of Instruction, University and Research (MIUR) through FFABR 2017 fund.
L.M. acknowledges support from the University of Rome Tor Vergata through "Mission: Sustainability 2016" fund.
The NGTS instrument and operations are funded by 
the University of Warwick,
the University of Leicester,
Queen's University Belfast,
the University of Geneva,
the Deutsches Zentrum f\" ur Luft- und Raumfahrt e.V. (DLR; under the `Gro\ss investition GI-NGTS'),
the University of Cambridge and the UK STFC (project reference ST/M001962/1). P.J.W.\ and R.G.W.\ acknowledge support by STFC through consolidated grants ST/L000733/1 and ST/P000495/1. 
ISB, OCW and RS achknowledge grants CNPQ (305737/2015-5, 312813/2013-9), CNPQ-PIBIC (37815/2016) \& FAPESP Proc. 2016/24561-0.%
TLK acknowledges use of the COAST facility, operated by the Open University.
%



\bibliographystyle{mnras}
\bibliography{references} 

\begin{thebibliography}{}
\makeatletter
\relax
\def\mn@urlcharsother{\let\do\@makeother \do\$\do\&\do\#\do\^\do\_\do\%\do\~}
\def\mn@doi{\begingroup\mn@urlcharsother \@ifnextchar [ {\mn@doi@}
  {\mn@doi@[]}}
\def\mn@doi@[#1]#2{\def\@tempa{#1}\ifx\@tempa\@empty \href
  {http://dx.doi.org/#2} {doi:#2}\else \href {http://dx.doi.org/#2} {#1}\fi
  \endgroup}
\def\mn@eprint#1#2{\mn@eprint@#1:#2::\@nil}
\def\mn@eprint@arXiv#1{\href {http://arxiv.org/abs/#1} {{\tt arXiv:#1}}}
\def\mn@eprint@dblp#1{\href {http://dblp.uni-trier.de/rec/bibtex/#1.xml}
  {dblp:#1}}
\def\mn@eprint@#1:#2:#3:#4\@nil{\def\@tempa {#1}\def\@tempb {#2}\def\@tempc
  {#3}\ifx \@tempc \@empty \let \@tempc \@tempb \let \@tempb \@tempa \fi \ifx
  \@tempb \@empty \def\@tempb {arXiv}\fi \@ifundefined
  {mn@eprint@\@tempb}{\@tempb:\@tempc}{\expandafter \expandafter \csname
  mn@eprint@\@tempb\endcsname \expandafter{\@tempc}}}

\bibitem[\protect\citeauthoryear{{Aizawa}, {Masuda}, {Kawahara}  \&
  {Suto}}{{Aizawa} et~al.}{2018}]{Aizawa18}
{Aizawa} M.,  {Masuda} K.,  {Kawahara} H.,   {Suto} Y.,  2018, \mn@doi [\aj]
  {10.3847/1538-3881/aab9a1}, \href
  {http://adsabs.harvard.edu/abs/2018AJ....155..206A} {155, 206}

\bibitem[\protect\citeauthoryear{{Ansdell} et~al.,}{{Ansdell}
  et~al.}{2016}]{Ansdell16a}
{Ansdell} M.,  et~al., 2016, \mn@doi [\apj] {10.3847/0004-637X/816/2/69}, \href
  {http://adsabs.harvard.edu/abs/2016ApJ...816...69A} {816, 69}

\bibitem[\protect\citeauthoryear{{Ansdell} et~al.,}{{Ansdell}
  et~al.}{2018}]{Ansdell18}
{Ansdell} M.,  et~al., 2018, \mn@doi [\mnras] {10.1093/mnras/stx2293}, \href
  {http://adsabs.harvard.edu/abs/2018MNRAS.473.1231A} {473, 1231}

\bibitem[\protect\citeauthoryear{{Armitage}}{{Armitage}}{2011}]{Armitage11}
{Armitage} P.~J.,  2011, \mn@doi [\araa] {10.1146/annurev-astro-081710-102521},
  \href {http://adsabs.harvard.edu/abs/2011ARA%26A..49..195A} {49, 195}

\bibitem[\protect\citeauthoryear{{Astropy Collaboration} et~al.,}{{Astropy
  Collaboration} et~al.}{2013}]{astropy:2013}
{Astropy Collaboration} et~al., 2013, \mn@doi [\aap]
  {10.1051/0004-6361/201322068}, \href
  {http://adsabs.harvard.edu/abs/2013A%26A...558A..33A …} {558, A33}

\bibitem[\protect\citeauthoryear{{Bertin} \& {Arnouts}}{{Bertin} \&
  {Arnouts}}{1996}]{1996A&AS..117..393B}
{Bertin} E.,  {Arnouts} S.,  1996, \mn@doi [\aaps] {10.1051/aas:1996164}, \href
  {http://adsabs.harvard.edu/abs/1996A%26AS..117..393B} {117, 393}

\bibitem[\protect\citeauthoryear{{Bouvier} et~al.,}{{Bouvier}
  et~al.}{1999}]{Bouvier99}
{Bouvier} J.,  et~al., 1999, \aap, \href
  {http://adsabs.harvard.edu/abs/1999A%26A...349..619B} {349, 619}

\bibitem[\protect\citeauthoryear{Bouvier et~al.,}{Bouvier
  et~al.}{2003}]{bouvier2003eclipses}
Bouvier J.,  et~al., 2003, Astronomy \& Astrophysics, 409, 169

\bibitem[\protect\citeauthoryear{Boyajian et~al.,}{Boyajian
  et~al.}{2016}]{boyajian2016planet}
Boyajian T.,  et~al., 2016, Monthly Notices of the Royal Astronomical Society,
  457, 3988

\bibitem[\protect\citeauthoryear{{Br{\"a}uer} \& {Fuhrmann}}{{Br{\"a}uer} \&
  {Fuhrmann}}{1992}]{1992Stern..68...19B}
{Br{\"a}uer} H.-J.,  {Fuhrmann} B.,  1992, Die Sterne, \href
  {http://adsabs.harvard.edu/abs/1992Stern..68...19B} {68, 19}

\bibitem[\protect\citeauthoryear{{Br{\"a}uer}, {H{\"a}usele}, {L{\"o}chel}  \&
  {Polko}}{{Br{\"a}uer} et~al.}{1999}]{1999AcHA....6...70B}
{Br{\"a}uer} H.-J.,  {H{\"a}usele} I.,  {L{\"o}chel} K.,   {Polko} N.,  1999,
  Acta Historica Astronomiae, \href
  {http://adsabs.harvard.edu/abs/1999AcHA....6...70B} {6, 70}

\bibitem[\protect\citeauthoryear{{Buchhave} et~al.,}{{Buchhave}
  et~al.}{2010}]{Buchhave:2010}
{Buchhave} L.~A.,  et~al., 2010, \mn@doi [\apj] {10.1088/0004-637X/720/2/1118},
  \href {http://adsabs.harvard.edu/abs/2010ApJ...720.1118B} {720, 1118}

\bibitem[\protect\citeauthoryear{Caballero}{Caballero}{2010}]{caballero2010occultation}
Caballero J.~A.,  2010, Astronomy \& Astrophysics, 511, L9

\bibitem[\protect\citeauthoryear{{Canup} \& {Ward}}{{Canup} \&
  {Ward}}{2002}]{Canup02}
{Canup} R.~M.,  {Ward} W.~R.,  2002, \mn@doi [\aj] {10.1086/344684}, \href
  {http://adsabs.harvard.edu/abs/2002AJ....124.3404C} {124, 3404}

\bibitem[\protect\citeauthoryear{Chote}{Chote}{2018}]{choteinprep}
Chote P.,  2018, "A Custom NGTS Pipeline" (in prep)

\bibitem[\protect\citeauthoryear{{Cody} \& {Hillenbrand}}{{Cody} \&
  {Hillenbrand}}{2014}]{Cody14}
{Cody} A.~M.,  {Hillenbrand} L.~A.,  2014, \mn@doi [\apj]
  {10.1088/0004-637X/796/2/129}, \href
  {http://adsabs.harvard.edu/abs/2014ApJ...796..129C} {796, 129}

\bibitem[\protect\citeauthoryear{{Collazzi}, {Schaefer}, {Xiao}, {Pagnotta},
  {Kroll}, {L{\"o}chel}  \& {Henden}}{{Collazzi}
  et~al.}{2009}]{2009AJ....138.1846C}
{Collazzi} A.~C.,  {Schaefer} B.~E.,  {Xiao} L.,  {Pagnotta} A.,  {Kroll} P.,
  {L{\"o}chel} K.,   {Henden} A.~A.,  2009, \mn@doi [\aj]
  {10.1088/0004-6256/138/6/1846}, \href
  {http://adsabs.harvard.edu/abs/2009AJ....138.1846C} {138, 1846}

\bibitem[\protect\citeauthoryear{{Colome} \& {Ribas}}{{Colome} \&
  {Ribas}}{2006}]{Colome2006}
{Colome} J.,  {Ribas} I.,  2006, IAU Special Session, \href
  {http://adsabs.harvard.edu/abs/2006IAUSS...6E..11C} {6, 11}

\bibitem[\protect\citeauthoryear{Coppejans et~al.,}{Coppejans
  et~al.}{2013}]{coppejans2013characterizing}
Coppejans R.,  et~al., 2013, Publications of the Astronomical Society of the
  Pacific, 125, 976

\bibitem[\protect\citeauthoryear{{DENIS Consortium}}{{DENIS
  Consortium}}{2005}]{2005yCat....102002D}
{DENIS Consortium} 2005, VizieR Online Data Catalog, \href
  {http://adsabs.harvard.edu/abs/2005yCat....102002D} {1}

\bibitem[\protect\citeauthoryear{{Dekker}, {D'Odorico}, {Kaufer}, {Delabre}  \&
  {Kotzlowski}}{{Dekker} et~al.}{2000}]{2000SPIE.4008..534D}
{Dekker} H.,  {D'Odorico} S.,  {Kaufer} A.,  {Delabre} B.,   {Kotzlowski} H.,
  2000, in {Iye} M.,  {Moorwood} A.~F.,  eds,  \procspie Vol. 4008, Optical and
  IR Telescope Instrumentation and Detectors. pp 534--545,
  \mn@doi{10.1117/12.395512}

\bibitem[\protect\citeauthoryear{{Donati}, {Semel}, {Carter}, {Rees}  \&
  {Collier Cameron}}{{Donati} et~al.}{1997}]{DonatiEtAl97}
{Donati} J.-F.,  {Semel} M.,  {Carter} B.~D.,  {Rees} D.~E.,   {Collier
  Cameron} A.,  1997, \mn@doi [\mnras] {10.1093/mnras/291.4.658}, \href
  {http://adsabs.harvard.edu/abs/1997MNRAS.291..658D} {291, 658}

\bibitem[\protect\citeauthoryear{{F{\H u}r{\'e}sz}}{{F{\H
  u}r{\'e}sz}}{2008}]{furesz:2008}
{F{\H u}r{\'e}sz} G.,  2008, PhD thesis, University of Szeged, Hungary

\bibitem[\protect\citeauthoryear{Fossey, Waldmann  \& Kipping}{Fossey
  et~al.}{2009}]{fossey2009detection}
Fossey S.~J.,  Waldmann I.~P.,   Kipping D.~M.,  2009, Monthly Notices of the
  Royal Astronomical Society: Letters, 396, L16

\bibitem[\protect\citeauthoryear{Ginski et~al.,}{Ginski
  et~al.}{2018}]{ginski2018first}
Ginski C.,  et~al., 2018, arXiv preprint arXiv:1805.02261

\bibitem[\protect\citeauthoryear{{Goranskij}, {Shugarov}, {Zharova}, {Kroll}
  \& {Barsukova}}{{Goranskij} et~al.}{2010}]{2010PZ.....30....4G}
{Goranskij} V.,  {Shugarov} S.,  {Zharova} A.,  {Kroll} P.,   {Barsukova}
  E.~A.,  2010, Peremennye Zvezdy, \href
  {http://adsabs.harvard.edu/abs/2010PZ.....30....4G} {30}

\bibitem[\protect\citeauthoryear{Hardy, Butterley, Dhillon, Littlefair  \&
  Wilson}{Hardy et~al.}{2015}]{hardy2015pt5m}
Hardy L.,  Butterley T.,  Dhillon V.,  Littlefair S.,   Wilson R.,  2015,
  Monthly Notices of the Royal Astronomical Society, 454, 4316

\bibitem[\protect\citeauthoryear{{Heising}, {Marcy}  \&
  {Schlichting}}{{Heising} et~al.}{2015}]{Heising15}
{Heising} M.~Z.,  {Marcy} G.~W.,   {Schlichting} H.~E.,  2015, \mn@doi [\apj]
  {10.1088/0004-637X/814/1/81}, \href
  {http://adsabs.harvard.edu/abs/2015ApJ...814...81H} {814, 81}

\bibitem[\protect\citeauthoryear{{Hippke} et~al.,}{{Hippke}
  et~al.}{2017}]{2017ApJ...837...85H}
{Hippke} M.,  et~al., 2017, \mn@doi [\apj] {10.3847/1538-4357/aa615d}, \href
  {http://adsabs.harvard.edu/abs/2017ApJ...837...85H} {837, 85}

\bibitem[\protect\citeauthoryear{Hunter}{Hunter}{2007}]{Hunter:2007}
Hunter J.~D.,  2007, \mn@doi [Computing In Science \& Engineering]
  {10.1109/MCSE.2007.55}, 9, 90

\bibitem[\protect\citeauthoryear{{Johnson}, {Schaefer}, {Kroll}  \&
  {Henden}}{{Johnson} et~al.}{2014}]{2014ApJ...780L..25J}
{Johnson} C.~B.,  {Schaefer} B.~E.,  {Kroll} P.,   {Henden} A.~A.,  2014,
  \mn@doi [\apjl] {10.1088/2041-8205/780/2/L25}, \href
  {http://adsabs.harvard.edu/abs/2014ApJ...780L..25J} {780, L25}

\bibitem[\protect\citeauthoryear{Kafka}{Kafka}{2016}]{kafka2016observations}
Kafka S.,  2016, Observations from the AAVSO International Database

\bibitem[\protect\citeauthoryear{Kennedy, Kenworthy, Pepper, Rodriguez, Siverd,
  Stassun  \& Wyatt}{Kennedy et~al.}{2017}]{kennedy2017transiting}
Kennedy G.~M.,  Kenworthy M.~A.,  Pepper J.,  Rodriguez J.~E.,  Siverd R.~J.,
  Stassun K.~G.,   Wyatt M.~C.,  2017, Royal Society open science, 4, 160652

\bibitem[\protect\citeauthoryear{Kenworthy \& Mamajek}{Kenworthy \&
  Mamajek}{2015}]{kenworthy2015modeling}
Kenworthy M.~A.,  Mamajek E.~E.,  2015, The Astrophysical Journal, 800, 126

\bibitem[\protect\citeauthoryear{{Kley} \& {Nelson}}{{Kley} \&
  {Nelson}}{2012}]{Kley12}
{Kley} W.,  {Nelson} R.~P.,  2012, \mn@doi [\araa]
  {10.1146/annurev-astro-081811-125523}, \href
  {http://adsabs.harvard.edu/abs/2012ARA%26A..50..211K} {50, 211}

\bibitem[\protect\citeauthoryear{Kochanek et~al.,}{Kochanek
  et~al.}{2017}]{kochanek2017all}
Kochanek C.,  et~al., 2017, Publications of the Astronomical Society of the
  Pacific, 129, 104502

\bibitem[\protect\citeauthoryear{{Lang}, {Hogg}, {Mierle}, {Blanton}  \&
  {Roweis}}{{Lang} et~al.}{2010}]{2010AJ....139.1782L}
{Lang} D.,  {Hogg} D.~W.,  {Mierle} K.,  {Blanton} M.,   {Roweis} S.,  2010,
  \mn@doi [\aj] {10.1088/0004-6256/139/5/1782}, \href
  {http://adsabs.harvard.edu/abs/2010AJ....139.1782L} {139, 1782}

\bibitem[\protect\citeauthoryear{{Magni} \& {Coradini}}{{Magni} \&
  {Coradini}}{2004}]{Magni04}
{Magni} G.,  {Coradini} A.,  2004, \mn@doi [\planss]
  {10.1016/j.pss.2003.08.028}, \href
  {http://adsabs.harvard.edu/abs/2004P%26SS...52..343M} {52, 343}

\bibitem[\protect\citeauthoryear{{Mallonn} et~al.,}{{Mallonn}
  et~al.}{2018}]{mallonn2018gj1214}
{Mallonn} M.,  et~al., 2018, \mn@doi [\aap] {10.1051/0004-6361/201732300},
  \href {http://adsabs.harvard.edu/abs/2018A%26A...614A..35M} {614, A35}

\bibitem[\protect\citeauthoryear{Mamajek, Quillen, Pecaut, Moolekamp, Scott,
  Kenworthy, Cameron  \& Parley}{Mamajek et~al.}{2012}]{mamajek2012planetary}
Mamajek E.~E.,  Quillen A.~C.,  Pecaut M.~J.,  Moolekamp F.,  Scott E.~L.,
  Kenworthy M.~A.,  Cameron A.~C.,   Parley N.~R.,  2012, The Astronomical
  Journal, 143, 72

\bibitem[\protect\citeauthoryear{McCormac, Pollacco, Skillen, Faedi, Todd  \&
  Watson}{McCormac et~al.}{2013}]{mccormac2013donuts}
McCormac J.,  Pollacco D.,  Skillen I.,  Faedi F.,  Todd I.,   Watson C.,
  2013, Publications of the Astronomical Society of the Pacific, 125, 548

\bibitem[\protect\citeauthoryear{McCormac, Skillen, Pollacco, Faedi, Ramsay,
  Dhillon, Todd  \& Gonzalez}{McCormac et~al.}{2014}]{mccormac2014search}
McCormac J.,  Skillen I.,  Pollacco D.,  Faedi F.,  Ramsay G.,  Dhillon V.,
  Todd I.,   Gonzalez A.,  2014, Monthly Notices of the Royal Astronomical
  Society, 438, 3383

\bibitem[\protect\citeauthoryear{{Nemmen}, {Georganopoulos}, {Guiriec},
  {Meyer}, {Gehrels}  \& {Sambruna}}{{Nemmen} et~al.}{2012}]{2012Nemmen}
{Nemmen} R.~S.,  {Georganopoulos} M.,  {Guiriec} S.,  {Meyer} E.~T.,  {Gehrels}
  N.,   {Sambruna} R.~M.,  2012, \mn@doi [Science] {10.1126/science.1227416},
  \href {http://adsabs.harvard.edu/abs/2012Sci...338.1445N} {338, 1445}

\bibitem[\protect\citeauthoryear{{Osborn} et~al.,}{{Osborn}
  et~al.}{2017}]{Osborn17}
{Osborn} H.~P.,  et~al., 2017, \mn@doi [\mnras] {10.1093/mnras/stx1249}, \href
  {http://adsabs.harvard.edu/abs/2017MNRAS.471..740O} {471, 740}

\bibitem[\protect\citeauthoryear{{Price-Whelan} et~al.,}{{Price-Whelan}
  et~al.}{2018}]{astropy:2018}
{Price-Whelan} A.~M.,  et~al., 2018, \mn@doi [\aj] {10.3847/1538-3881/aabc4f},
  \href {https://ui.adsabs.harvard.edu/#abs/2018AJ....156..123T …} {156, 123}

\bibitem[\protect\citeauthoryear{Ricker et~al.,}{Ricker
  et~al.}{2010}]{ricker2010transiting}
Ricker G.~R.,  et~al., 2010, in Bulletin of the American Astronomical Society.
  p.~459

\bibitem[\protect\citeauthoryear{Rodriguez, Pepper, Stassun, Siverd, Cargile,
  Beatty  \& Gaudi}{Rodriguez et~al.}{2013}]{rodriguez2013occultation}
Rodriguez J.~E.,  Pepper J.,  Stassun K.~G.,  Siverd R.~J.,  Cargile P.,
  Beatty T.~G.,   Gaudi B.~S.,  2013, The Astronomical Journal, 146, 112

\bibitem[\protect\citeauthoryear{{Rodriguez} et~al.,}{{Rodriguez}
  et~al.}{2015}]{Rodriguez2015}
{Rodriguez} J.~E.,  et~al., 2015, \mn@doi [\aj] {10.1088/0004-6256/150/1/32},
  \href {http://adsabs.harvard.edu/abs/2015AJ....150...32R} {150, 32}

\bibitem[\protect\citeauthoryear{{Ryabchikova}, {Piskunov}, {Kurucz},
  {Stempels}, {Heiter}, {Pakhomov}  \& {Barklem}}{{Ryabchikova}
  et~al.}{2015}]{VALD3}
{Ryabchikova} T.,  {Piskunov} N.,  {Kurucz} R.~L.,  {Stempels} H.~C.,  {Heiter}
  U.,  {Pakhomov} Y.,   {Barklem} P.~S.,  2015, \mn@doi [\physscr]
  {10.1088/0031-8949/90/5/054005}, \href
  {http://adsabs.harvard.edu/abs/2015PhyS...90e4005R} {90, 054005}

\bibitem[\protect\citeauthoryear{Scaringi et~al.,}{Scaringi
  et~al.}{2016}]{scaringi2016peculiar}
Scaringi S.,  et~al., 2016, Monthly Notices of the Royal Astronomical Society,
  463, 2265

\bibitem[\protect\citeauthoryear{Shappee et~al.,}{Shappee
  et~al.}{2014}]{shappee2014all}
Shappee B.,  et~al., 2014, in American Astronomical Society Meeting Abstracts\#
  223.

\bibitem[\protect\citeauthoryear{{Smette} et~al.,}{{Smette}
  et~al.}{2015}]{SmetteEtAl15}
{Smette} A.,  et~al., 2015, \mn@doi [\aap] {10.1051/0004-6361/201423932}, \href
  {https://ui.adsabs.harvard.edu/#abs/2015A&A...576A..77S} {576, A77}

\bibitem[\protect\citeauthoryear{{Southworth} et~al.,}{{Southworth}
  et~al.}{2009}]{southworth09mn}
{Southworth} J.,  et~al., 2009, MNRAS, \href {2009MNRAS.396.1023S} {396, 1023}

\bibitem[\protect\citeauthoryear{{Southworth} et~al.,}{{Southworth}
  et~al.}{2014}]{southworth14mn}
{Southworth} J.,  et~al., 2014, MNRAS, \href {2014MNRAS.444..776S} {444, 776}

\bibitem[\protect\citeauthoryear{Strassmeier et~al.,}{Strassmeier
  et~al.}{2004}]{strassmeier2004stella}
Strassmeier K.,  et~al., 2004, Astronomische Nachrichten, 325, 527

\bibitem[\protect\citeauthoryear{{Teachey}, {Kipping}  \& {Schmitt}}{{Teachey}
  et~al.}{2018}]{Kipping18}
{Teachey} A.,  {Kipping} D.~M.,   {Schmitt} A.~R.,  2018, \mn@doi [\aj]
  {10.3847/1538-3881/aa93f2}, \href
  {http://adsabs.harvard.edu/abs/2018AJ....155...36T} {155, 36}

\bibitem[\protect\citeauthoryear{Vanderburg, Rappaport  \& Mayo}{Vanderburg
  et~al.}{2018}]{vanderburg2018detecting}
Vanderburg A.,  Rappaport S.~A.,   Mayo A.~W.,  2018, arXiv preprint
  arXiv:1805.01903

\bibitem[\protect\citeauthoryear{{Waagen}}{{Waagen}}{2017}]{waagen2017}
{Waagen} E.~O.,  2017, Alert Notice 584: Monitoring of PDS 110 requested to
  cover upcoming eclipse by exoplanet, \url
  {https://www.aavso.org/aavso-alert-notice-584}

\bibitem[\protect\citeauthoryear{{Ward} \& {Canup}}{{Ward} \&
  {Canup}}{2010}]{Ward10}
{Ward} W.~R.,  {Canup} R.~M.,  2010, \mn@doi [\aj]
  {10.1088/0004-6256/140/5/1168}, \href
  {http://adsabs.harvard.edu/abs/2010AJ....140.1168W} {140, 1168}

\bibitem[\protect\citeauthoryear{{Watson}, {Dhillon}  \& {Shahbaz}}{{Watson}
  et~al.}{2006}]{WatsonEtAl06}
{Watson} C.~A.,  {Dhillon} V.~S.,   {Shahbaz} T.,  2006, \mn@doi [\mnras]
  {10.1111/j.1365-2966.2006.10130.x}, \href
  {http://adsabs.harvard.edu/abs/2006MNRAS.368..637W} {368, 637}

\bibitem[\protect\citeauthoryear{Wheatley et~al.,}{Wheatley
  et~al.}{2018}]{wheatley2017next}
Wheatley P.~J.,  et~al., 2018, Monthly Notices of the Royal Astronomical
  Society, 475, 4476

\bibitem[\protect\citeauthoryear{{Zacharias}, {Urban}, {Zacharias}, {Wycoff},
  {Hall}, {Germain}, {Holdenried}  \& {Winter}}{{Zacharias}
  et~al.}{2003}]{2003yCat.1289....0Z}
{Zacharias} N.,  {Urban} S.~E.,  {Zacharias} M.~I.,  {Wycoff} G.~L.,  {Hall}
  D.~M.,  {Germain} M.~E.,  {Holdenried} E.~R.,   {Winter} L.,  2003, VizieR
  Online Data Catalog, \href
  {http://adsabs.harvard.edu/abs/2003yCat.1289....0Z} {1289}

\bibitem[\protect\citeauthoryear{{de Mooij}, {Watson}  \& {Kenworthy}}{{de
  Mooij} et~al.}{2017}]{DeMooijEtAl17}
{de Mooij} E.~J.~W.,  {Watson} C.~A.,   {Kenworthy} M.~A.,  2017, \mn@doi
  [\mnras] {10.1093/mnras/stx2142}, \href
  {http://adsabs.harvard.edu/abs/2017MNRAS.472.2713D} {472, 2713}

\bibitem[\protect\citeauthoryear{{de Mooij} et~al.}{{de Mooij}
  et~al.}{prep}]{DeMooijEtAlPrep}
{de Mooij} E.~J.~W.,  et~al., in prep.

\makeatother
\end{thebibliography}

\section*{Affiliations}
{\it
$^{\ddagger}$~AAVSO contributor\\
$^1$~Aix Marseille Universit\'e, CNRS, LAM (Laboratoire d'Astrophysique de Marseille) UMR 7326, F-13388, Marseille, France\\ 
$^2$~Leiden Observatory, Leiden University, P.O. Box 9513, 2300 RA Leiden, The Netherlands\\
$^3$~Harvard-Smithsonian Center for Astrophysics, 60 Garden St, Cambridge, MA 02138, USA\\
$^4$~Department of Physics, University of Warwick, Gibbet Hill Road, Coventry, CV4 7AL, UK\\
$^5$~Centre for Exoplanets and Habitability, University of Warwick, Gibbet Hill Road, Coventry, CV4 7AL, UK\\
$^6$~Las Cumbres Observatory, 6740 Cortona Drive, Suite 102, Goleta, CA  93117, USA\\
$^7$~School of Physical Sciences, and Centre for Astrophysics and Relativity, Dublin City University, Glasnevin, Dublin 9, Ireland\\
$^8$~Sonneberg Observatory, Sternwartestr. 32, 96515 Sonneberg, Germany\\
$^{9}$~ UNESP-S\~a Paulo State University, Grupo de Din\^amica Orbital e Planetologia, CEP 12516-410 Guaratinguet\'a, SP, Brazil\\
$^{10}$~Acton Sky Portal (Private observatory), Acton, MA, USA.\\
$^{11}$~Astronomical Observatory, Dipartimento di Scienze Fisiche, della Terra e dell'Ambiente, University of Siena, Italy\\
$^{12}$~Department of Physics and Astronomy, Leicester Institute of Space and Earth Observation, University of Leicester, LE1 7RH, UK \\
$^{13}$~Centre for Advanced Instrumentation, Department of Physics, University of Durham, South Road, Durham DH1 3LE, UK\\
$^{14}$~Department of Astronomy, Stockholm University, Alba Nova University Center, SE-106 91, Stockholm, Sweden\\
$^{15}$~Dept. of Physics \& Astronomy, Univ. of Sheffield, Sheffield, S3 7RH\\
$^{16}$~Instituto de Astrof\'{\i}sica de Canarias, E-38205 La Laguna, Tenerife, Spain\\
$^{17}$~astroLAB IRIS, Verbrandemolenstraat 5, 8902 Zillebeke, Belgium\\
$^{18}$~Keele University Astrophysics Group Newcastle-under-Lyme ST5 5BG UK\\
$^{19}$~UCL Observatory (UCLO), 553 Watford Way, Mill Hill, London NW7 2QS\\
$^{20}$~Dept. of Physics and Astronomy, University College London, Gower St., London WC1E 6BT\\
$^{21}$~Astrophysics Group, Cavendish Laboratory, J.J. Thomson Avenue, Cambridge CB3 0HE, UK\\ 
$^{22}$~Vereniging Voor Sterrenkunde (VVS), Brugge, BE-8000, Belgium\\
$^{23}$~Montsec Astronomical Observatory (OAdM), Institut d'Estudis Espacials de Catalunya (IEEC), Barcelona, Spain\\
$^{24}$~Physics \& Astronomy, York University, Toronto, Ontario L3T 3R1, Canada\\
$^{25}$~American Association of Variable Star Observers,49 Bay State Road, Cambridge, MA 02138, USA\\
$^{26}$~Leibniz-Institut f{\"u}r Astrophysik Potsdam (AIP), An der Sternwarte 16, D-14482 Potsdam, Germany\\
$^{27}$~Department of Physics, University of Rome Tor Vergata, Via della Ricerca Scientifica 1, I-00133 -- Roma, Italy \\
$^{28}$~Max-Planck-Institut f{\"u}r Astronomie K{\"o}nigstuhl 17. D-69117 Heidelberg Germany\\
$^{29}$~INAF -- Astrophysical Observatory of Turin, Via Osservatorio 20, I-10025 -- Pino Torinese, Italy \\
$^{30}$~Medical University of Bialystok, Faculty of Medicine, 15-089 Bialystok, Poland\\
$^{31}$~Department of Physics and Astronomy, Shumen University, 9700 Shumen, Bulgaria\\
$^{32}$~Arkansas Tech University 1701 N. Boulder Ave. Russellville, AR 72801-2222\\
$^{33}$~SUPA, School of Physics \& Astronomy, North Haugh, St Andrews, KY16 9SS, United Kingdom\\
$^{34}$~Institute of Planetary Research, German Aerospace Center, Rutherfordstr. 2, 12489 Berlin, Germany\\
$^{35}$~Perth Exoplanet Survey Telescope (PEST), Australia\\
$^{36}$~Astrophysics Research Centre, Queen's University Belfast, Belfast, BT7 1NN, UK\\
}





\appendix

\section{Observer Offsets}

\onecolumn
\begin{table}
	\centering
	\caption{Information for each source of $BVRI$ photometry during the 2017 observing campaign. Offsets are in relative flux. They are sorted by number of exposures, although this is not necessarily a proxy for photometric quality or observation duration. LCOGT data was re-adjusted such that the median matches the archive value in each band. $\dagger$ denotes those values held fixed. $\star$ demarks where data was initially binned. "OTHER" denotes AAVSO observers with fewer than 25 observations.}
	\label{tab:photometry}
	\resizebox{\textwidth}{!}{
       \begin{tabular}{lcccccccc} 
		\hline
		Observatory & $N_{\rm{img}}(B)$ & B Offset & $N_{\rm{img}}(V)$ & V Offset & $N_{\rm{img}}(R)$ & R Offset & $N_{\rm{img}}(I)$ & I Offset\\
		\hline
 LCOGT 0.4m &  --- & --- &  --- & --- &  --- & --- &  --- & --- \\ 
NITES & 230 &  $ -0.0129 \pm 0.001 $  & 211 &  $ -0.0074^{+0.0038}_{-0.0018} $  & 202 &  $ -0.0105 \pm 0.0008 $  & 202 &  $ 0.0018^{+0.0012}_{-0.0017} $  \\ 
LCOGT 1m & 215 &  $ 0^{\dagger} $  & 196 &  $ 0^{\dagger} $  & 205 &  $ 0^{\dagger} $  &  --- & --- \\ 
STELLA & 134 &  $ -0.0081 \pm 0.0011 $  & 125 &  $ -0.0067^{+0.0013}_{-0.0017} $  &  --- & --- & 131 &  $ -0.0006 \pm 0.0012 $  \\ 
NGTS &  --- & --- &  --- & --- &  --- & --- &  --- & --- \\ 
CAHA & 60 &  $ -1.2297^{+0.0023}_{-0.0031} $  & 65 &  $ -0.703 \pm 0.0037 $  & 63 &  $ -3.1034^{+0.0024}_{-0.0033} $  & 63 &  $ 0^{\dagger} $  \\ 
ASAS-SN &  --- & --- & 237 &  $ -0.0104 \pm 0.0011 $  &  --- & --- &  --- & --- \\ 
FEG &  --- & --- & 137 &  $ 0.0493^{+0.0027}_{-0.0013} $  &  --- & --- &  --- & --- \\ 
TJO & 51 &  $ -0.083^{+0.0025}_{-0.002} $  &  --- & --- &  --- & --- & 45 &  $ 0.532^{+0.04}_{-0.076} $  \\ 
pt5m & 18 &  $ -0.0914^{+0.0052}_{-0.0037} $  & 22 &  $ -0.004^{+0.003}_{-0.004} $  &  --- & --- &  --- & --- \\ 
SAAO &  --- & --- & 11$^{\star}$ &  $ 0.0228 \pm 0.0074 $  & 10$^{\star}$ &  $ -0.4365 \pm 0.0018 $  & 6$^{\star}$ &  $ -0.2683 \pm 0.0011 $  \\ 
UCLO &  --- & --- &  --- & --- & 9$^{\star}$ &  $ -0.0021^{+0.0026}_{-0.0055} $  & 9$^{\star}$ &  $ 0.0063^{+0.0018}_{-0.0038} $  \\ 
\hline
AAVSO/LCLC &  --- & --- & 1898 &  $ -0.00054^{+0.00092}_{-0.00097} $  &  --- & --- &  --- & --- \\ 
AAVSO/QULA & 365 &  $ -0.0603^{+0.0019}_{-0.0012} $  & 266 &  $ 0.0001^{+0.0023}_{-0.0026} $  & 338 &  $ -0.3945^{+0.0021}_{-0.0031} $  & 500 &  $ -0.2745^{+0.0038}_{-0.0022} $  \\ 
AAVSO/MGW & 374 &  $ -0.0253 \pm 0.0011 $  & 369 &  $ 0.02156^{+0.00078}_{-0.00052} $  & 353 &  $ -0.37578^{+0.0007}_{-0.00082} $  & 366 &  $ -0.2183^{+0.0018}_{-0.0011} $  \\ 
AAVSO/HMB & 455 &  $ -0.0451 \pm 0.001 $  & 555 &  $ 0.0125 \pm 0.001 $  &  --- & --- & 439 &  $ -0.2786^{+0.0064}_{-0.0037} $  \\ 
AAVSO/JM & 325 &  $ -0.073 \pm 0.012 $  & 329 &  $ 0.047 \pm 0.008 $  &  --- & --- &  --- & --- \\ 
AAVSO/RJWA &  --- & --- &  --- & --- &  --- & --- &  --- & --- \\ 
AAVSO/DLM &  --- & --- & 281 &  $ 0.00703^{+0.00074}_{-0.00078} $  &  --- & --- &  --- & --- \\ 
AAVSO/HKEB & 73 &  $ -0.0179 \pm 0.0024 $  & 73 &  $ -0.0065 \pm 0.002 $  & 76 &  $ -0.354^{+0.003}_{-0.002} $  &  --- & --- \\ 
AAVSO/PVEA & 73 &  $ -0.05 \pm 0.0019 $  & 58 &  $ 0.0135^{+0.0013}_{-0.0022} $  &  --- & --- & 58 &  $ -0.2334 \pm 0.0014 $  \\ 
AAVSO/MXI & 47 &  $ -0.0676 \pm 0.0023 $  & 42 &  $ 0.017 \pm 0.002 $  & 40 &  $ -0.3967 \pm 0.0021 $  & 39 &  $ -0.267 \pm 0.0008 $  \\ 
AAVSO/BSM &  --- & --- & 83 &  $ -0.0213^{+0.0012}_{-0.0022} $  &  --- & --- & 78 &  $ -0.3358^{+0.0057}_{-0.0027} $  \\ 
AAVSO/HJW & 33 &  $ -0.0725^{+0.0076}_{-0.0055} $  & 87 &  $ 0.0055^{+0.003}_{-0.0038} $  &  --- & --- & 33 &  $ -0.2736 \pm 0.0021 $  \\ 
AAVSO/BPAD & 34 &  $ -0.0492^{+0.0033}_{-0.0044} $  & 44 &  $ 0.0056^{+0.0015}_{-0.0015} $  & 39 &  $ -0.4245 \pm 0.002 $  & 35 &  $ -0.2893 \pm 0.0017 $  \\ 
AAVSO/DERA &  --- & --- & 56 &  $ -0.0051 \pm 0.003 $  &  --- & --- & 57 &  $ -0.2534^{+0.0071}_{-0.0046} $  \\ 
AAVSO/LMA &  --- & --- & 97 &  $ 0.0185^{+0.0025}_{-0.0022} $  &  --- & --- &  --- & --- \\ 
AAVSO/RLUB &  --- & --- & 90 &  $ 0.0158^{+0.002}_{-0.0026} $  &  --- & --- &  --- & --- \\ 
AAVSO/FSTC &  --- & --- & 83 &  $ 0.0113^{+0.0017}_{-0.0053} $  &  --- & --- &  --- & --- \\ 
AAVSO/LPAC & 29 &  $ -0.0477^{+0.002}_{-0.0015} $  & 41 &  $ -0.0182 \pm 0.0032 $  &  --- & --- &  --- & --- \\ 
AAVSO/KCLA & 18 &  $ -0.0544^{+0.0045}_{-0.0031} $  & 17 &  $ 0.0001 \pm 0.002 $  & 17 &  $ -0.4125^{+0.0025}_{-0.0014} $  & 17 &  $ -0.2825^{+0.0052}_{-0.0016} $  \\ 
AAVSO/DKS & 25 &  $ -0.0182^{+0.0039}_{-0.0046} $  & 25 &  $ 0.014^{+0.0037}_{-0.0026} $  &  --- & --- &  --- & --- \\ 
AAVSO/BLOC & 10 &  $ -0.149^{+0.0043}_{-0.0067} $  & 20 &  $ -0.0231^{+0.0039}_{-0.0032} $  & 10 &  $ -0.4063^{+0.0054}_{-0.007} $  &  --- & --- \\ 
AAVSO/OTHER & 3 &  $ -0.0676^{+0.0018}_{-0.004} $  & 30 &  $ 0.0266^{+0.0031}_{-0.0071} $  & 5 &  $ -0.3999^{+0.0032}_{-0.0055} $  &  --- & --- \\ 
AAVSO/TTG & 9 &  $ -0.0608^{+0.0012}_{-0.0022} $  & 8 &  $ -0.0006^{+0.0013}_{-0.0017} $  & 9 &  $ -0.3846^{+0.0016}_{-0.0013} $  & 12 &  $ -0.2479^{+0.0014}_{-0.0015} $  \\ 
\hline
	\end{tabular}
}
\end{table}

\begin{table}
	\centering
	\caption{Information for each source of $ugriz$ photometry during the 2017 observing campaign. LCOGT data was re-adjusted such that the median matches the archive value in each band.}
	\label{tab:photometry_ugriz}
	\begin{tabular}{lcccccccccc} 
		\hline
		Observatory & $N_{\rm{img}}(u)$ & u Offset & $N_{\rm{img}}(g)$ & g Offset & $N_{\rm{img}}(r)$ & r Offset & $N_{\rm{img}}(i)$ & i Offset & $N_{\rm{img}}(z)$ & z Offset \\
\hline
LCOGT 0.4m &  --- & --- & 405 &  $ 0^{\dagger} $  & 402 &  $ 0^{\dagger} $  & 380 &  $ 0^{\dagger} $  & 378 &  $ 0^{\dagger} $  \\ 
LCOGT 1m & 204 &  $ 0^{\dagger} $  &  --- & --- &  --- & --- &  --- & --- &  --- & --- \\ 
pt5m &  --- & --- &  --- & --- & 23 &  $ -0.1771 \pm 0.0076 $  &  --- & --- &  --- & --- \\ 
\hline
AAVSO/RJWA &  --- & --- & 164 &  $ -0.9615^{+0.0059}_{-0.0094} $  & 192 &  $ -0.2654^{+0.0034}_{-0.0042} $  &  --- & --- &  --- & --- \\ 
\hline
	\end{tabular}
\end{table}

\bsp	
\label{lastpage}

\twocolumn









\end{document}